# Calculation of core-excited and core-ionized states using variational quantum deflation method and applications to photocatalyst modelling

Soichi Shirai,[†]* Takahiro Horiba[†] and Hirotoshi Hirai[†]

[†] Toyota Central R&D Laboratories, Inc., Nagakute, Aichi 480−1192, Japan

* Author to whom correspondence should be addressed.

E-mail: shirai@mosk.tytlabs.co.jp

**ABSTRACT**

The possibility of performing quantum chemical calculations using quantum computers has attracted much interest. In this regard, variational quantum deflation (VQD) is a quantum-classical hybrid algorithm for the calculation of excited states with noisy intermediate-scale quantum (NISQ) devices. Although the validity of this method has been demonstrated, there have been few practical applications, primarily because of the uncertain effect of calculation conditions on the results. In the present study, calculations of the core-excited and core-ionized states for common molecules based on the VQD method were simulated using a classical computer, focusing on the effects of the weighting coefficients applied in the penalty terms of the cost function. Adopting a simplified procedure for estimating the weighting coefficients based on molecular orbital levels allowed these core-level states to be successfully calculated. The O $1s$ core-ionized state for a water molecule was calculated with various weighting coefficients and the resulting ansatz states were systematically examined. The application of this technique to functional materials was demonstrated by calculating the core-level states for titanium dioxide ($TiO_2$) and nitrogen-doped $TiO_2$ models. The results demonstrate that VQD calculations employing an appropriate cost function can be applied to the analysis of functional materials in conjunction with an experimental approach.

## 1. Introduction

The concept of performing quantum chemical calculations using quantum computers has attracted significant attention.[1,2] In this regard, the configuration interaction (CI) framework is the most common means of assessing electron correlation. In this framework, the wavefunction is described as a linear combination of Slater determinants or spin-adapted configuration state functions that represent the electron configurations.[3] If all the electron configurations generated from combinations of all the electrons and orbitals of the system are included, the method is termed full CI.[3,4] Although exact wavefunctions within an adopted basis set can be obtained using full CI calculations, the computational cost increases quite rapidly as the number of orbitals and electrons increases, and can easily outgrow the capacity of standard computers. In contrast, quantum computers could potentially perform the required calculations in polynomial time using the quantum phase estimation (QPE) argorithm.[5–8] Unfortunately, the QPE method requires fault tolerance and cannot be practically executed on current quantum computers (so-called noisy intermediate-scale quantum (NISQ) devices) that operate on the basis of qubits without error correction.[9–11] Accordingly, a quantum-classical hybrid algorithm (termed the variational quantum eigensolver (VQE)) has emerged as a means of allowing quantum chemical calculations to be performed using NISQ devices.[12,13]

Following the development of the VQE method to compute the ground state for a given system, VQE-based algorithms for excited states have also been proposed, including quantum subspace expansion (QSE),[14] multiscale-contracted VQE (MCVQE),[15] subspace-search VQE (SSVQE),[16] quantum equation of motion VQE (qEOM-VQE)[17] and others. The variational quantum deflation (VQD) method introduced by Higgott et al. is one such algorithm and allows computation of excited states based on the ground state obtained using the VQE method.[18] The superior efficiency and accuracy of the VQD approach compared with the SSVQE and MCVQE methods have been demonstrated based on calculations of the excited states for lithium hydride, diazene and azobenzene.[19] The VQD method has also been used to investigate the excited states for phenylsulfonylcarbazole compounds during the development of thermally activated delayed fluorescence (TADF) emitters for organic light emitting diode (OLED) applications.[20] These studies suggest that the VQD method could be employed in the research and development of functional materials such as light emitting devices,[21] photocatalysts,[22] photovoltanic cells and[23] photochromic materials.[24]

In VQD calculations, the excited states are sequentially computed in conjunction with minimization of the defined cost function.[18] Including penalty terms within the cost function is the most general means of ensuring that the computed wavefunction will be orthogonal to other states and have the desired eigenvalues.[13,18,25,26] Penalty terms are introduced to raise the cost in the case that ansatz states are non-orthogonal to other states, or eigenvalues deviate from user-specified values. Thus, the weighting coefficients that determine the magnitudes of the penalty terms dictate whether or not the desired excited states have been obtained, and so these coefficients must be carefully selected. Even so, the coefficients are typically chosen heuristically. Although small coefficients are preferable so as to obtain rapid convergence, undesired excited states can possibly be obtained if the coefficients are too small. Conversely, if the coefficients are too large, the optimization may not converge properly. To address this problem, Kuroiwa and Nakagawa proposed a convenient formula to estimate the weighting coefficients as a part of their systematic analysis of the penalty terms.[27] Assuming that this formula is generally applicable to calculations of the excited states for common molecules, it could greatly increase the practical applications of the VQD method. Although this formula was initially assessed based on calculations of simple model systems such as $H_2$ and $H_4$, it should be examined to determine if it is more widely applicable to the excited states for more complex molecules. A practical procedure for utilizing this method should also be explored.

On this basis, the present study applied the VQD method to calculate the core-excited and core-ionized states for common molecules based on simulated quantum circuit computations with a classical computer. The absorption bands derived from the core-excited state calculations were experimentally verified using X-ray absorption near edge spectroscopy (XANES),[28–31] while the core-ionization energies were confirmed using X-ray photoelectron spectroscopy (XPS).[32–35] It should be noted that the core-ionization energy is also referred to as the core-electron binding energy (CEBE) and that XPS is also traditionally known as electron spectroscopy for chemical analysis (ESCA). These spectroscopic techniques are widely utilized for the analysis of molecular structures and to determine the surface compositions of condensed materials, local bonding environments of specific atomic species and electronic properties such as oxidation states. Computations of core-level states are of interest with regard to fundamental scientific studies and also with respect to the theoretical analyses of spectra.[36,37] If the calculation of core-level states using the VQD method is found to be possible, this technique could have practical applications in conjunction with experimental analyses. As part of this study, the effects of the weighting coefficients used in the penalty terms were systematically analyzed by calculating the O $1s$ core-ionized state for a water molecule. The core-excitation energies and core-ionization energies for this molecule are on the order of hundreds of eV or higher, and so are much larger than those of the more well-studied valence excited states. Therefore, the correlation of the weighting coefficients with the results could be clearly visualized. As examples of applications to functional materials, the O $1s$ core-excited states and Ti $2p$ core-ionized states for titanium dioxide ($TiO_2$)[38–41] and nitrogen-doped $TiO_2$ (N-$TiO_2$)[42,43] were also calculated. These materials were selected because anatase $TiO_2$ is well-known for its photocatalytic activity and N-$TiO_2$ is a visible-light-sensitive photocatalyst.

Note that singly core-excited and singly core-ionized states were calculated in this study. Because these states have less multiconfiguration character, their wavefunctions can be reasonably described using a small number of electron configurations. Thus, even though only a small number of qubits can be currently employed in simulations on classical computers (as with calculations using actual NISQ devices), the excitation and ionization energies in such cases can be evaluated with high accuracy so that the calculation results can be verified by comparison with experimental data.[44] It is noteworthy that Bauman at al. recently calculated the doubly excited core-level states for a water molecule using the QPE algorithm.[45]

## 2. Theory

2.1. VQD method and weighting coefficients of penalty terms

Here, the VQD algorithm is briefly reviewed[18] as well as the penalty terms related to the cost function.[27] In the VQD algorithm, the $k$-th ansatz state, $|\psi(\theta_k)\rangle$, is calculated by minimizing the cost function

$$L(\theta_k) = \langle\psi(\theta_k)|\hat{H}|\psi(\theta_k)\rangle + \sum_{i=0}^{k-1} \beta\langle\psi(\theta_k)|\psi(\theta_i)\rangle + L_{penalty}. \quad (1)$$

Here $\hat{H}$ is a Hamiltonian written as $\hat{H} = \sum c_j \widehat{P_j}$, where $\widehat{P_j}$ is a single-qubit Pauli operator and $c_j$ is the corresponding coefficient. We denote the eigenstates up to the $(k-1)$th as $|\psi(\theta_i)\rangle$ ($i = 0, 1, \ldots, k-1$) and assume that $|\psi(\theta_i)\rangle$ has been obtained before the calculation of $|\psi(\theta_k)\rangle$. The ground state, $|\psi(\theta_0)\rangle$, is initially obtained using the VQE method, after which the $|\psi(\theta_1)\rangle$ state is calculated. In a similar manner, the $|\psi(\theta_2)\rangle$, $|\psi(\theta_3)\rangle$, … and $|\psi(\theta_k)\rangle$ states are calculated one by one in a sequential fashion. The second term of Eq. (1) is added to ensure that $|\psi(\theta_k)\rangle$ is orthogonal to $|\psi(\theta_i)\rangle$ while $\beta$ is a weighting coefficient referred to as the overlap weight. Unless $|\psi(\theta_k)\rangle$ is orthogonal to one of the $|\psi(\theta_i)\rangle$ states, $L(\theta_k)$ is increased by $\beta$ because of this second term. The third term, $L_{penalty}$, is introduced to impose a constraint on the spin multiplicity, spin quantum number and number of electrons associated with $|\psi(\theta_k)\rangle$. The $L_{penalty}$ adopted in this study consisted of three terms, written as

$$L_{penalty} = w_1 \left\langle\psi(\theta_k)\left|\left(\hat{S}^2 - o_1\right)^2\right|\psi(\theta_k)\right\rangle + w_2 \left\langle\psi(\theta_k)\left|\left(\widehat{S_z} - o_2\right)^2\right|\psi(\theta_k)\right\rangle + w_3 \left\langle\psi(\theta_k)\left|\left(\widehat{S_N} - o_3\right)^2\right|\psi(\theta_k)\right\rangle, \quad (2)$$

where $\hat{S}^2$ is the total spin-squared operator, $\widehat{S_z}$ is the $z$-component of the total spin operator, $\hat{S}$, $\widehat{S_N}$ is the particle-number operator, $w_1$, $w_2$ and $w_3$ are weighting coefficients termed the s2 number weight, sz number weight and particle number weight, respectively, and $o_1$, $o_2$ and $o_3$ are the eigenvalues of $\hat{S}^2$, $\widehat{S_z}$ and $\widehat{S_N}$, respectively, that must be satisfied by $|\psi(\theta_k)\rangle$. Here, the $o_x$ values ($x = 1$–$3$) are input parameters. $L_{penalty}$ goes to zero when the eigenvalues of $|\psi(\theta_k)\rangle$ coincide with the user-specified $o_x$ values. In contrast, if the eigenvalues of $|\psi(\theta_k)\rangle$ deviate from the $o_x$ values, $L(\theta_k)$ increases because $L_{penalty}$ is nonzero. When the operators in Eq. (2) act on $|\psi(\theta_k)\rangle$, we have

$$L_{penalty} = w_1\{S(S+1) - o_1\}^2 + w_2(M_S - o_2)^2 + w_3(N - o_3)^2, \tag{3}$$

where $S$ is the total spin angular momentum, $M_S$ is the spin quantum number, and $N$ is the number of electrons. If we define $\Delta_x$ as the deviation of the eigenvalues of $|\psi(\theta_k)\rangle$ from $o_x$, we obtain

$$L_{penalty} = w_1(\Delta_1)^2 + w_2(\Delta_2)^2 + w_3(\Delta_3)^2, \tag{4}$$

where $\Delta_1 = S(S+1) - o_1$, $\Delta_2 = M_S - o_2$, and $\Delta_3 = N - o_3$.

2.2. Estimation of weighting coefficients for excited state calculations

2.2.1. Estimation of $\beta$

If $|\psi(\theta_k)\rangle$ is orthogonal to $|\psi(\theta_i)\rangle$ and all the $\Delta_x$ values are zero, $L(\theta_k)$ is equal to the energy of $|\psi(\theta_k)\rangle$, meaning that

$$L(\theta_k) = \langle\psi(\theta_k)|\widehat{H}|\psi(\theta_k)\rangle = E(\theta_k). \tag{5}$$

Next, we assume an ansatz state $|\psi(\theta_k^*)\rangle$ that is non-orthogonal to one of the $|\psi(\theta_i)\rangle$ states. Consequently, the $\Delta_x$ values of $|\psi(\theta_k^*)\rangle$ are zero and its cost function, $L(\theta_k^*)$, can be written as

$$L(\theta_k^*) = \langle\psi(\theta_k^*)|\hat{H}|\psi(\theta_k^*)\rangle + \sum_{i=0}^{k-1}\beta\langle\psi(\theta_k^*)|\psi(\theta_i)\rangle = E(\theta_k^*) + \beta. \tag{6}$$

If $L(\theta_k) > L(\theta_k^*)$, $|\psi(\theta_k^*)\rangle$ is obtained instead of $|\psi(\theta_k)\rangle$. To obtain $|\psi(\theta_k)\rangle$, $\beta$ must satisfy the condition

$$\beta > E(\theta_k) - E(\theta_k^*), \tag{7}$$

so that $L(\theta_k^*) > L(\theta_k)$. The right-hand side of Eq. (7) is the energy gap between $|\psi(\theta_k)\rangle$ and $|\psi(\theta_k^*)\rangle$, which is maximized if $|\psi(\theta_k^*)\rangle$ is the ground state: $|\psi(\theta_0)\rangle$. We then have the condition for $\beta$ written as

$$\beta > E(\theta_k) - E(\theta_0). \tag{8}$$

This condition was provided in the original paper in which VQD was described.[18] Unfortunately, the right-hand side of Eq. (8) is the excitation energy of $|\psi(\theta_k)\rangle$, which is a part of the calculation results. As a practical alternative, the excitation energy of the singly excited state can be estimated from the energy gap between the occupied and virtual orbitals related to the excitation, as

$$\beta > \varepsilon_{kv} - \varepsilon_{ko}, \tag{9}$$

where $\varepsilon_{ko}$ and $\varepsilon_{kv}$ are the energy levels of the occupied and virtual orbitals that are relevant to the main configuration of $|\psi(\theta_k)\rangle$, respectively.

In this study, the $\beta$ values were estimated based on Eq. (9), while the $\varepsilon_{ko}$ and $\varepsilon_{kv}$ values were obtained through the Hartree-Fock calculation that was used to obtain an initial state for the ground state

VQE calculation. In general, the gap between $\varepsilon_{kv}$ and $\varepsilon_{ko}$ will be larger than that between $E(\theta_k)$ and $E(\theta_0)$ when determined using the Hartree-Fock method.[46] Consequently, the $\beta$ value estimated from Eq. (9) should satisfy Eq. (8). Note that the value of $\varepsilon_{kv} - \varepsilon_{ko}$ based on the use of Kohn-Sham orbitals may be smaller than $E(\theta_k) - E(\theta_0)$. In addition, calculations using the density functional theory (DFT) method adopting an exchange-correlation functional tend to give a smaller energy gap between the occupied and virtual orbitals compared with the corresponding excitation energy.[46]

2.2.2. Estimation of $w_1$

The values of $w_x$ ($x = 1 - 3$) were estimated using the formula proposed by Kuroiwa and Nakagawa.[27] These calculations assumed $|\psi(\theta_k^*)\rangle$ with $\Delta_p \neq 0$ and $\Delta_q = \Delta_r = 0$, where $(p, q, r) \in x$, $p \neq q$, $p \neq r$ and $q \neq r$. In addition, $|\psi(\theta_k^*)\rangle$ was orthogonal to $|\psi(\theta_i)\rangle$ because $\Delta_p \neq 0$. In this case, $L(\theta_k^*)$ could be written as

$$L(\theta_k^*) = \langle \psi(\theta_k^*)|\hat{H}|\psi(\theta_k^*)\rangle + w_p(\Delta_p)^2 = E(\theta_k^*) + w_p(\Delta_p)^2. \tag{10}$$

According to Eqs. (5) and (10), to have $L(\theta_k^*) > L(\theta_k)$,

$$w_p > \frac{E(\theta_k) - E(\theta_k^*)}{(\Delta_p)^2} \tag{11}$$

must be true. Therefore, the lower limit of $w_p$ is obtained when $|\Delta_p|$ is at a minimum, $|\Delta_p^{min}|$, such that

$$w_p > \frac{E(\theta_k) - E(\theta_k^*)}{(\Delta_p^{min})^2}, \tag{12}$$

where $\Delta_p^{min}$ is the minimum deviation of the eigenvalue from the user-specified $o_p$. Equation (12)

indicates that the lower limit of $w_p$ can be estimated based on $E(\theta_k) - E(\theta_k^*)$ and $\Delta_p^{min}$.

We also assumed that, in the case of $p = 1$, $|\psi(\theta_k^*)\rangle$ has $\Delta_1 \neq 0$ and $\Delta_2 = \Delta_3 = 0$ and $|\psi(\theta_k^*)\rangle$ is orthogonal to $|\psi(\theta_i)\rangle$ when $\Delta_1 \neq 0$. This gives a $L(\theta_k^*)$ value equal to

$$L(\theta_k^*) = \langle\psi(\theta_k^*)|\hat{H}|\psi(\theta_k^*)\rangle + w_1(\Delta_1)^2 = E(\theta_k^*) + w_1(\Delta_1)^2. \tag{13}$$

Because the eigenvalue of $\hat{S}^2$ is $S(S+1)$ and the minimum change in $S$ is 1/2, $|\Delta_1^{min}|$ is 3/4. Hence, according to Eq. (12),

$$w_1 > \frac{16}{9}\left(E(\theta_k) - E(\theta_k^*)\right), \tag{14}$$

should be true. The right-hand side of Eq. (14) is maximized when $E(\theta_k^*) = E(\theta_0)$, such that

$$w_1 > \frac{16}{9}\left(E(\theta_k) - E(\theta_0)\right). \tag{15}$$

The energy gap between $E(\theta_k)$ and $E(\theta_0)$ is the excitation energy for $E(\theta_k)$, which can be approximated based on $\varepsilon_{kv} - \varepsilon_{ko}$ in a similar manner to Eq. (9) as

$$w_1 > \frac{16}{9}(\varepsilon_{kv} - \varepsilon_{ko}). \tag{16}$$

In the present work, Eq. (16) was adopted as the condition for $w_1$.

2.2.3. Estimation of $w_2$

We assumed the case of $p = 2$, meaning $|\psi(\theta_k^*)\rangle$ with $\Delta_2 \neq 0$ and $\Delta_1 = \Delta_3 = 0$ and $|\psi(\theta_k^*)\rangle$ orthogonal to $|\psi(\theta_i)\rangle$ when $\Delta_2 \neq 0$. $L(\theta_k^*)$ in this case is

$$L(\theta_k^*) = \langle\psi(\theta_k^*)|\hat{H}|\psi(\theta_k^*)\rangle + w_2(\Delta_2)^2 = E(\theta_k^*) + w_2(\Delta_2)^2. \tag{17}$$

Since the eigenvalue of $\widehat{S}_z$ is $M_S$ and the minimum change in $M_S$ is ±1/2, $|\Delta_2^{min}|$ is 1/2. According to Eq. (12), the $w_2$ value should satisfy $w_2 > 4\bigl(E(\theta_k) - E(\theta_k^*)\bigr)$ so that $L(\theta_k^*) > L(\theta_k)$. The gap between $E(\theta_k)$ and $E(\theta_k^*)$ is maximized when $E(\theta_k^*) = E(\theta_0)$, and the gap between $E(\theta_k)$ and $E(\theta_0)$ can be approximated by $\varepsilon_{kv} - \varepsilon_{ko}$ as with Eq. (9), such that

$$w_2 > 4(\varepsilon_{kv} - \varepsilon_{ko}). \tag{18}$$

The $w_2$ values in this study were estimated based on Eq. (18).

2.2.4. Estimation of $w_3$

We assumed the case of $p = 3$, meaning $|\psi(\theta_k^*)\rangle$ with $\Delta_3 \neq 0$ and $\Delta_1 = \Delta_2 = 0$ and $|\psi(\theta_k^*)\rangle$ orthogonal to $|\psi(\theta_i)\rangle$ when $\Delta_3 \neq 0$. $L(\theta_k^*)$ in this case is

$$L(\theta_k^*) = \langle\psi(\theta_k^*)|\hat{H}|\psi(\theta_k^*)\rangle + w_3(\Delta_3)^2 = E(\theta_k^*) + w_3(\Delta_3)^2. \tag{19}$$

Because the eigenvalue of $\widehat{S}_N$ is the number of electrons, its minimum change is ±1 and, accordingly, $|\Delta_3^{min}|$ is 1. Therefore, the $w_3$ value should satisfy $w_3 > E(\theta_k) - E(\theta_k^*)$ so that $L(\theta_k^*) > L(\theta_k)$. The gap between $E(\theta_k)$ and $E(\theta_k^*)$ is maximized when $E(\theta_k^*) = E(\theta_0)$, and the gap between $E(\theta_k)$ and $E(\theta_0)$ can be approximated by $\varepsilon_{kv} - \varepsilon_{ko}$. Thus, we have

$$w_3 > \varepsilon_{kv} - \varepsilon_{ko}. \tag{20}$$

The $w_3$ values in this study were estimated based on Eq. (20). Overall, the weighting coefficients for the calculations of core-excited states were estimated so that

$$\beta = \frac{9}{16}w_1 = \frac{1}{4}w_2 = w_3 > \varepsilon_{kv} - \varepsilon_{ko}. \tag{21}$$

2.3. Estimation of weighting coefficients for core-ionized state calculations

In the present study, each core-ionized state was calculated as a high-energy excited state for a molecule in its ionized state. In this procedure, the valence-ionized state was obtained as $|\psi(\theta_0)\rangle$ via the VQE calculations. Accordingly, the charge-neutral ground state having a lower energy than $|\psi(\theta_0)\rangle$ could be obtained as $|\psi(\theta_k^*)\rangle$ and $L(\theta_k^*)$ in this case was

$$L(\theta_k^*) = \langle\psi(\theta_k^*)|\hat{H}|\psi(\theta_k^*)\rangle + L_{penalty}. \tag{22}$$

If $|\psi(\theta_k)\rangle$ is the core-ionized state, $L_{penalty}$ should satisfy the condition

$$L_{penalty} > \langle\psi(\theta_k)|\hat{H}|\psi(\theta_k)\rangle - \langle\psi(\theta_k^*)|\hat{H}|\psi(\theta_k^*)\rangle, \tag{23}$$

such that $L(\theta_k^*) > L(\theta_k)$.

The right-hand side of Eq. (23) is the core-ionization energy resulting from the VQD calculation. This energy can be approximated as the negative of the core-orbital energy ($\varepsilon_{ko}$) according to Koopman's theorem[47]

$$L_{penalty} > -\varepsilon_{ko}. \tag{24}$$

Equation (24) indicates that the weighting coefficients used in calculations of core-ionized states should be estimated based on $-\varepsilon_{ko}$ instead of $\varepsilon_{kv} - \varepsilon_{ko}$, in contrast to the calculations for the excited states. Accordingly, this work adopted the condition

$$\beta = \frac{9}{16} w_1 = \frac{1}{4} w_2 = w_3 > -\varepsilon_{ko}, \tag{25}$$

and the $\varepsilon_{ko}$ values obtained from the Hartree-Fock calculations were used.

## 3. Computational Details

Optimizations of molecular geometries via coupled cluster singles, doubles and perturbative triples (CCSD(T))[48–50] calculations and DFT calculations with the B3LYP functional[51,52] were carried out using the Gaussian 09 program.[53] The VQE and VQD calculations with the optimized structures were performed using the Qamuy program version 0.26.1,[54] in conjunction with a hardware-efficient ansatz.[55] The Broyden-Fletcher-Goldfarb-Shanno (BFGS) algorithm was used to update variational parameters on a classical computer[56] while the Jordan-Wigner transformation was used to map Hamiltonians into a qubit circuit.[57–58] The depth of the simulated quantum circuit ($D$) was given a value of 10. Noise was not considered with regard to the simulated quantum circuit, meaning that exact expected values were used in the simulations.

Because twice as many qubits as the number of active orbitals were required within the framework detailed above, the computational cost of the simulated calculations on a classical computer increased rapidly with increasing number of active orbitals. Accordingly, the active space of the VQD calculations had to be constructed using as few active orbitals as possible. In addition, the molecular orbitals were greatly perturbed by core hole formation. A core hole attracts electrons, leading to radial contraction of the outer orbitals, such that the core-hole state becomes highly stable. This effect is referred to as orbital relaxation, and the effects of orbital relaxation were not fully incorporated into the CI calculations using a limited number of electron configurations. In such cases, orbital optimization can be an effective means of improving the accuracy of the calculations[44] and so VQD calculations with and without molecular orbital optimization were compared. The calculations with and without orbital optimization corresponded to conventional complete active space self-consistent field (CASSCF)[59,60] and complete active space configuration interaction (CASCI) calculations,[61] respectively. A flexible basis set was also required to deal with the significant changes of the orbitals and Dunning's cc-VXZ basis set (X = D, T, Q) was adopted.[62–64] The weighting coefficients were estimated based on the Hartree-Fock orbital energies with the cc-pVDZ basis set.

3.1. Calculations of core-excited states for CO, H$_2$CO and HCN

The $1s \rightarrow \pi^*$ core-excited states for CO, H$_2$CO and HCN molecules were calculated. The molecular geometries were optimized using the CCSD(T)/cc-pVQZ approach and the core-excited states were calculated using the VQD method with the cc-pVXZ basis set. The $1s \rightarrow \pi^*$ core-excitation energies were obtained as the energy gaps between the ground and core-excited states. In the case of CO and H$_2$CO, both the C $1s$ and O $1s$ core-excited states were calculated. Similarly, both the C $1s$ and N $1s$ core-excited states were calculated for HCN. The $1s$ and $\pi^*$ orbitals were selected as the active orbitals and two electrons in these orbitals were treated as active. In the following discussion, the CAS constructed from these active orbitals and electrons is denoted as CAS(2e, 2o). During the calculations for CO and HCN, two $1s \rightarrow \pi^*$ core-excited states were found to be degenerate because two $\pi^*$ orbitals had the same energy level, and so only one state was calculated. For H$_2$CO, the $\pi^*$ orbitals were split into different energy values because of the lower symmetry of the molecule. The weighting coefficients were estimated from the energy gaps between the $1s$ and $\pi^*$ orbitals in conjunction with Eq. (21).

3.2. Calculations of core-ionized states for CH$_4$, NH$_3$, H$_2$O and FH

The $1s$ core-ionized states for CH$_4$, NH$_3$, H$_2$O and FH molecules were calculated. The molecular geometries were optimized using the CCSD(T)/cc-pVQZ method and the core-excited states were calculated using the VQD method with the cc-pVXZ basis set. The $1s$ core-ionization energies were obtained as the energy gaps between the charge-neutral (closed-shell) singlet ground state and doublet core-ionized state. The $1s$ orbital and the highest occupied orbital (HOMO) were selected as the active orbitals and three electrons in these orbitals were treated as active electrons. Henceforth, the CAS constructed from these active orbitals and electrons is denoted as CAS(3e, 2o). When adopting CAS(3e, 2o) as the configuration space, the valence ionized state having the configuration $(1s)^2(HOMO)^1$ was the ground state and the core-ionized state was the excited state having the primary configuration $(1s)^1(HOMO)^2$. Accordingly, the core-ionized states could be calculated using the VQD algorithm. In a full CI calculation using a quantum computer with a larger number of qubits, all the occupied and virtual orbitals would be treated as active orbitals, such that the core-ionized states would be obtained as the excited states. The calculations presented herein based on CAS(3e, 2o) simulated this type of full CI calculation. The weighting coefficients for the VQD calculations were estimated based on the $1s$ orbital energies and using Eq. (25).

### 3.3. Effects of the weighting coefficients

The relationship between the weighting coefficients and the ansatz states was examined by calculating the $1s$ core-ionized state for a $H_2O$ molecule while varying the weighting coefficients. The molecular structure was optimized via a CCSD(T)/cc-pVQZ calculation while VQD calculations were carried out using the cc-pVDZ basis set.

### 3.4. Application to the $TiO_2$ and N-$TiO_2$ models

#### 3.4.1. Preparation of the models

In the quantum chemical calculations of solid materials such as metals and metal oxides, a small cluster of atoms is often employed as a model.[65] The calculation results obtained using this model cluster approach are significantly affected by the size and shape of the cluster, as well as the manner in which dangling bonds are terminated, all of which are arbitrarily assigned. Nakatsuji et al. proposed three principles for the construction of clusters: neutrality, stoichiometry and coordination.[66] In this work, we adopted a (TiO$_2$)$_{16}$ cluster, which is known to represent the smallest cluster of anatase-type TiO$_2$ that satisfies these three conditions.[67] Specifically, this cluster is charge neutral, has a stoichiometric Ti:O ratio of 1:2 (equal to that of the bulk system) and has no dangling bonds. The coordination numbers of the titanium atoms in this cluster were four or more, while those of the oxygen atoms were two or three. We initially prepared a (TiO$_2$)$_{16}$ cluster and optimized the geometrical structure using the B3LYP/cc-pVDZ basis set. Following this, a smaller cluster with one titanium atom and four oxygen atoms was cut out from the surface of this (TiO$_2$)$_{16}$ cluster and its dangling bonds were terminated with hydrogen atoms to obtain Ti(OH)$_4$. The positions of the four hydrogen atoms in the Ti(OH)$_4$ were reoptimized using the B3LYP/cc-pVDZ basis set, holding the other atoms at fixed positions during the re-optimization. Finally, one of the oxygen atoms of Ti(OH)$_4$ was replaced with a nitrogen atom and the newly-generated dangling bond on the nitrogen atom was saturated with a hydrogen atom to obtain Ti(OH)$_3$(NH$_2$). The positions of the hydrogen atoms of Ti(OH)$_3$(NH$_2$) were re-optimized using the B3LYP/cc-pVDZ basis set with the positions of the other atoms fixed. The resulting Ti(OH)$_4$ and Ti(OH)$_3$(NH$_2$) clusters were employed as models of TiO$_2$ and N-TiO$_2$, respectively.

3.4.2. Calculation of valence excited states, O 1$s$ core-excited states and Ti 2$p$ core-ionized states

The HOMO → LUMO excited states, O 1$s$ → LUMO core-excited states and Ti 2$p$ core-ionized states for Ti(OH)$_4$ and Ti(OH)$_3$NH$_2$ were calculated utilizing the VQD algorithm. The HOMO and LUMO of a solid material correspond to the valence band maximum (VBM) and the conduction band minimum (CBM), respectively. Therefore, absorption band edge shifts can be estimated from HOMO → LUMO excitation energies. The peak with the lowest energy in the O K-edge XANES spectrum was derived from the O 1$s$ → LUMO excitation, and the corresponding peak in the XPS spectrum was situated at the Ti 2$p$ core-ionization energy. The effects of nitrogen doping on these spectroscopic features could be examined based on calculations involving the Ti(OH)$_4$ and Ti(OH)$_3$(NH$_2$) cluster models.

The CAS for the calculations of the HOMO → LUMO excited states was constructed using these two orbitals and the two electrons in the HOMO. Similarly, the CAS for the calculations of the O 1$s$ → LUMO excited states was constructed using these two orbitals and the two electrons in the O 1$s$ orbital. The CAS adopted for these calculations is denoted herein as CAS(2e, 2o). During the calculations of the Ti 2$p$ core-ionized states, one of the Ti 2$p$ orbitals and the HOMO were selected as the active orbitals and three electrons in these orbitals were treated as active electrons. This CAS is denoted as CAS(3e, 2o). Since there are three Ti 2$p$ orbitals, three VQD calculations were made for each model, employing a different Ti 2$p$ orbital as an active orbital.

The energy levels of the Ti 2$p$ core-ionized states are split into $2p_{1/2}$ and $2p_{3/2}$ levels because of spin-orbit interactions.[68] Unfortunately, the ab initio algorithm for relativistic effects is not presently implemented in the Qamuy program, and so this splitting had to be approximated. Although the gap between the $2p_{1/2}$ and $2p_{3/2}$ states ($\Delta_{\text{SO}}$) of Ti varies depending on the material, the value is typically within the narrow range of 5.60 – 6.13 eV[69–71] and so the $\Delta_{\text{SO}}$ value was assumed to be 5.7 eV in the present work.[70] In the case of the core-ionized doublet state for an atom, the $2p_{1/2}$ energy level will be lower by $(3/2)\Delta_{\text{SO}}$ than the 2$p$ level, while the $2p_{3/2}$ energy level will be higher by $(1/2)\Delta_{\text{SO}}$ (Figure S1). Consequently, the core-ionization energies of Ti(OH)$_4$ and Ti(OH)$_3$NH$_2$, $E_{2p1/2}$ and $E_{2p3/2}$ derived from the $2p_{1/2}$ and $2p_{3/2}$ states, respectively, were estimated as

$$E_{2p1/2} = E_{2p} + \frac{2}{3}\Delta_{\text{SO}}, \tag{26}$$

and

$$E_{2p3/2} = E_{2p} - \frac{1}{3}\Delta_{\text{SO}}, \tag{27}$$

where $E_{2p}$ is the Ti $2p$ core-ionization energy calculated using the VQD method.

It should be noted that calculations using cluster models provide values for gas phase molecules, while the experimental values for TiO₂ and N-TiO₂ have been obtained using solid materials. When determining the ionization energies of finite systems such as molecules, the vacuum level is usually chosen as the reference. In contrast, the ionization energies of solids and surfaces are typically reported relative to the Fermi levels.[72–74] Thus, the standard for the ionization energy will be different between molecules in the gas phase and solid materials. In addition, a core hole in a solid can be greatly stabilized by charge redistributions in the surrounding continuum, while such effects are absent in the gas phase.[75,76] Therefore, the core-ionization energies that were calculated based on the cluster models could not be directly compared to the experimental data obtained from solids. For this reason, we also adjusted the calculated values so that the simulated $E_{2p}$ value for Ti(OH)₄ coincided with the experimental $E_{2p}$ value for bulk TiO₂ as estimated from the $E_{2p1/2}$ and $E_{2p3/2}$ values. It is noteworthy that theoretical methods for the calculation of absolute core ionization energies in solids from first principles have been investigated in detail.[73–77]

## 4. Results and Discussion

### 4.1. Calculations of core-excited states for CO, H₂CO and HCN

The $1s$ ($\varepsilon_{1s}$) and $\pi^*$ ($\varepsilon_{\pi*}$) orbital energies for CO, H₂CO and HCN molecules and the weighting coefficients used in the VQD calculations are summarized in Table 1. Note that the $\varepsilon_{\pi*} - \varepsilon_{1s}$ value was rounded up to an integer prior to being used for the estimation of the weighting coefficients. Because the $\varepsilon_{1s}$ energies increased in the order of C $1s$ > N $1s$ > O $1s$, the weighting coefficients were in the reverse order of O $1s$ > N $1s$ > C $1s$. The calculated core-excitation energies are presented in Table 2 together with available experimental values.[78–81] The $1s \rightarrow \pi^*$ core-excited states could be successfully calculated using the VQD method with the weighting coefficients in Table 1, and the calculated core-excitation energies deviated from the experimental values by only 3 – 5% in the case that Hartree-Fock orbitals were employed. Remarkably, these deviations were further reduced to less than 1.1% by incorporating orbital relaxations. In particular, discrepancies of less than 0.4% were achieved using the cc-pVTZ or cc-pVQZ basis set, meaning that the core-excitation energies calculated using the VQD method were in quantitatively good agreement with the experimental data. The CAS contained only the minimum configurations describing the wavefunction of the core-excited state, and so the results were highly dependent on whether or not orbital optimizations were applied. The calculated core-excitation energies of $H_2CO$ were lower than those of CO as well as the experimental values. The calculated core-excitation energies were also correlated with the local chemical-bonding states, indicating that the VQD calculations could be applied to structural analyses based on comparisons with experimental XANES spectra.

The calculated core-excitation energies were reduced depending on the basis set in the order of cc-pVDZ > cc-pVTZ > cc-pVQZ when orbital optimizations were carried out. A larger basis set was advantageous because it allowed orbital relaxation effects on core hole formation to be incorporated, thus reducing the energy of the core-excited state relative to that of the ground state. This obvious effect of the basis set size on the core-excitation energy was not observed in calculations performed without orbital optimization. Thus, the effects of enlarging the basis set were also fully incorporated in association with orbital optimization. Nevertheless, the calculated core-excitation energies deviated from the experimental values even when the larger cc-pVTZ and cc-pVQZ basis sets were used. The remaining discrepancies can be attributed to the lack of dynamical correlation within the present calculations.

**Table 1.** Core 1s orbital energies ($\varepsilon_{1s}$), $\pi^*$ orbital energies ($\varepsilon_{\pi*}$) obtained using the Hartree-Fock method, their differences ($\varepsilon_{\pi*} - \varepsilon_{1s}$) and weighting coefficients in the cost function for the VQD calculations: overlap weights ($\beta$), s2 number weights ($w_1$), sz number weights ($w_2$) and particle number weights ($w_3$). All values are in hartrees.

| Molecule | Energy of core orbital | | $\varepsilon_{\pi*}$ | $\varepsilon_{\pi*} - \varepsilon_{1s}$ | $\beta$ | $w_1$ | $w_2$ | $w_3$ |
|---|---|---|---|---|---|---|---|---|
| | Orbital | $\varepsilon_{1s}$ | | | | | | |
| CO | C 1s | −11.3612 | 0.1276 | 11.4888 | 12.0 | 21.4 | 48.0 | 12.0 |
| | O 1s | −20.6649 | 0.1276 | 20.7925 | 21.0 | 37.4 | 84.0 | 21.0 |
| H$_2$CO | C 1s | −11.3451 | 0.1356 | 11.4807 | 12.0 | 21.4 | 48.0 | 12.0 |
| | O 1s | −20.5757 | 0.1356 | 20.7113 | 21.0 | 37.4 | 84.0 | 21.0 |
| HCN | C 1s | −11.2919 | 0.1515 | 11.4434 | 12.0 | 21.4 | 48.0 | 12.0 |
| | N 1s | −15.5989 | 0.1515 | 15.7504 | 16.0 | 28.5 | 64.0 | 16.0 |

**Table 2.** Calculated $1s \rightarrow \pi^*$ core excitation energies for CO, H$_2$CO and HCN. Available experimental values are also shown. All energies are in eV and the values in parentheses are deviations from the experimental values.

| Molecule | Core orbital | | Basis set | Core excitation energy | | | |
|---|---|---|---|---|---|---|---|
| | | | | Without orbital optimization | | With orbital optimization | |
| CO | C 1s | calc. | cc-pVDZ | 298.39 | (+3.82%) | 290.47 | (+1.07%) |
| | | | cc-pVTZ | 298.79 | (+3.96%) | 288.39 | (+0.34%) |
| | | | cc-pVQZ | 299.50 | (+4.21%) | 288.10 | (+0.24%) |
| | | exptl.[a] | | | | 287.4 | |
| | O 1s | calc. | cc-pVDZ | 552.92 | (+3.52%) | 536.14 | (+0.38%) |
| | | | cc-pVTZ | 553.25 | (+3.59%) | 533.97 | (−0.03%) |
| | | | cc-pVQZ | 553.74 | (+3.68%) | 533.57 | (−0.10%) |
| | | exptl.[b] | | | | 534.1 | |
| H$_2$CO | C 1s | calc. | cc-pVDZ | 297.99 | (+4.34%) | 288.54 | (+1.03%) |
| | | | cc-pVTZ | 298.20 | (+4.41%) | 286.58 | (+0.34%) |
| | | | cc-pVQZ | 298.73 | (+4.60%) | 286.37 | (+0.27%) |
| | | exptl.[c] | | | | 285.6 | |
| | O 1s | calc. | cc-pVDZ | 548.83 | (+3.40%) | 532.95 | (+0.41%) |
| | | | cc-pVTZ | 549.35 | (+3.50%) | 530.82 | (+0.00%) |
| | | | cc-pVQZ | 549.88 | (+3.60%) | 530.45 | (−0.07%) |
| | | exptl.[c] | | | | 530.8 | |
| HCN | C 1s | calc. | cc-pVDZ | 299.36 | (+4.53%) | 289.38 | (+1.04%) |
| | | | cc-pVTZ | 299.55 | (+4.59%) | 287.27 | (+0.30%) |
| | | | cc-pVQZ | 300.15 | (+4.80%) | 287.01 | (+0.21%) |
| | | exptl.[d] | | | | 286.4 | |
| | N 1s | calc. | cc-pVDZ | 415.33 | (+3.91%) | 402.53 | (+0.71%) |
| | | | cc-pVTZ | 415.78 | (+4.02%) | 400.28 | (+0.15%) |
| | | | cc-pVQZ | 416.49 | (+4.20%) | 399.97 | (+0.07%) |
| | | exptl.[d] | | | | 399.7 | |

[a] Reference 78. [b] Reference 79. [c] Reference 80. [d] Reference 81.

### 4.2. Calculations of core-ionized states for CH$_4$, NH$_3$, H$_2$O and FH

The 1$s$ orbital energies for CH$_4$, NH$_3$, H$_2$O and FH molecules and the weighting coefficients used in the VQD calculations are summarized in Table 3. The $\varepsilon_{1s}$ values were rounded up to the nearest integer when used for the estimation of weighting coefficients. The weighting coefficients decreased in the order of FH > H$_2$O > NH$_3$ > CH$_4$ because the $\varepsilon_{1s}$ energies were in the order of C 1$s$ > N 1$s$ > O 1$s$ > F 1$s$. The calculated core-ionization energies are collected in Table 4 along with available experimental data.[82] By adopting the weighting coefficients in Table 3, the 1$s$ core-ionized states were successfully calculated as $|\psi(\theta_1)\rangle$. The resulting energies were found to deviate from the experimental values by 3.01 – 4.24% when orbital optimization was not applied, although these discrepancies were significantly reduced (to less than 1%) by carrying out orbital optimizations. The core-ionization energies obtained using the VQD method were in quantitatively good agreement with the experimental data in the case that orbital optimizations were employed, as was also the case for the core-excited state calculations. These results indicate that the peak position in an XPS spectrum can be predicted with reasonable accuracy using VQD calculations.

The calculated core-ionization energies decreased with increasing size of the adopted basis set when applying orbital optimization, in the order of cc-pVDZ > cc-pVTZ > cc-pVQZ. This ordering suggests that orbital relaxation, along with core-hole formation, could be more accurately described using a more flexible basis set, leading to a downward shift of the energy level of the core-ionized state relative to that of the neutral ground state.[44] The core-ionization energies calculated without orbital optimization were not clearly correlated with the size of the basis set because the effects of the orbital relaxations were only insufficiently considered. The core-ionization energies were found to be generally underestimated in the case that larger basis sets such as the cc-pVTZ or cc-pVQZ sets were employed. It has also been reported that the core-ionization energy is typically underestimated in the framework of the ΔSCF method.[83] Therefore, the deviations from the experimental values can be attributed to a lack of incorporated dynamical correlation. The charge-neutral ground state has one more electron than the core-ionized state, and so it is reasonable to expect that the extent of dynamical electron correlation in the charge-neutral ground state will be larger than that in the core-ionized state. Therefore, the gap between these states is underestimated when the dynamical correlation is not fully considered.

**Table 3.** Core 1s orbital energies ($\varepsilon_{1s}$) obtained from Hartree-Fock calculations and weighting coefficients in the cost function for the VQD calculations: overlap weights ($\beta$), s2 number weights ($w_1$), sz number weights ($w_2$) and particle number weights ($w_3$). All values are in hartrees.

| Molecule | $\varepsilon_{1s}$ | $\beta$ | $w_1$ | $w_2$ | $w_3$ |
|---|---|---|---|---|---|
| $CH_4$ | −11.2152 | 12.0 | 21.4 | 48.0 | 12.0 |
| $NH_3$ | −15.5367 | 16.0 | 28.5 | 64.0 | 16.0 |
| $H_2O$ | −20.5508 | 21.0 | 37.4 | 84.0 | 21.0 |
| FH | −26.2781 | 27.0 | 48.0 | 108.0 | 27.0 |

**Table 4.** Calculated core ionization energies for $CH_4$, $NH_3$, $H_2O$ and FH. Available experimental values are also included. All energies are in eV and values in parentheses are percentage deviations from the experimental data.

| Molecule | Calc. or exptl. | Basis set | Core ionization energy | | | |
|---|---|---|---|---|---|---|
| | | | Without orbital optimization | | With orbital optimization | |
| $CH_4$ | calc. | cc-pVDZ | 302.23 | (+3.91%) | 290.51 | (−0.12%) |
| | | cc-pVTZ | 302.05 | (+3.85%) | 288.35 | (−0.86%) |
| | | cc-pVQZ | 302.15 | (+3.88%) | 288.20 | (−0.92%) |
| | exptl.[a] | | | | 290.86 | |
| $NH_3$ | calc. | cc-pVDZ | 422.77 | (+4.24%) | 407.63 | (+0.51%) |
| | | cc-pVTZ | 422.70 | (+4.22%) | 405.64 | (+0.02%) |
| | | cc-pVQZ | 422.78 | (+4.24%) | 405.41 | (−0.04%) |
| | exptl.[a] | | | | 405.57 | |
| $H_2O$ | calc. | cc-pVDZ | 559.21 | (+3.58%) | 541.64 | (+0.33%) |
| | | cc-pVTZ | 559.32 | (+3.61%) | 539.60 | (−0.05%) |
| | | cc-pVQZ | 559.46 | (+3.63%) | 539.27 | (−0.11%) |
| | exptl.[a] | | | | 539.86 | |
| FH | calc. | cc-pVDZ | 715.05 | (+3.01%) | 695.84 | (+0.24%) |
| | | cc-pVTZ | 715.28 | (+3.04%) | 693.63 | (−0.08%) |
| | | cc-pVQZ | 715.41 | (+3.06%) | 693.20 | (−0.14%) |
| | exptl.[a] | | | | 694.18 | |

[a] Reference 83.

### 4.3. Effect of the weighting coefficients

The present calculations were carried out while varying the weighting coefficients, and Figure 1 shows the primary configurations of the electronic states obtained for the H$_2$O molecule. The calculated electronic state energies, eigenvalues, and imposed penalties are provided in Table 5. Table 6 summarizes the calculated energies and the assignments of the first, second, and third electronic states: $|\psi(\theta_0)\rangle$, $|\psi(\theta_1)\rangle$ and $|\psi(\theta_2)\rangle$. The costs of these states are summarized in Table 7. Below, we review the results with respect to each coefficient.

With regard to condition (i), the core-ionized state represented by $|12\bar{2}\rangle$ was obtained as the ansatz state $|\psi(\theta_1)\rangle$ with $\beta \geq 21.0$. In contrast, the state was $|\psi(\theta_2)\rangle$ and the ground state $|1\bar{1}2\rangle$ was obtained as both $|\psi(\theta_0)\rangle$ and $|\psi(\theta_1)\rangle$ with $\beta < 21.0$. In the case of $\beta < 21.0$, $|\psi(\theta_1)\rangle$ was non-orthogonal to $|\psi(\theta_0)\rangle$ and the penalty of $\beta$ was imposed on its cost function. However, $\langle 1\bar{1}2|\hat{H}|1\bar{1}2\rangle + \beta$ was still less than the cost function for the core-ionized state $|12\bar{2}\rangle$. Therefore, the ground state $|1\bar{1}2\rangle$ was obtained again as $|\psi(\theta_1)\rangle$. If $|\psi(\theta_2)\rangle$ is also $|1\bar{1}2\rangle$, its cost function will be $\langle 1\bar{1}2|\hat{H}|1\bar{1}2\rangle + 2\beta$ because $|\psi(\theta_2)\rangle$ is non-orthogonal to both $|\psi(\theta_0)\rangle$ and $|\psi(\theta_1)\rangle$, and so the cost function will be greater than that of the core-ionized state $|12\bar{2}\rangle$. Thus, the core-ionized state $|12\bar{2}\rangle$ was obtained as $|\psi(\theta_2)\rangle$. In the case with $\beta \geq 21.0$, the cost function for $|1\bar{1}2\rangle$ became greater than that for $|12\bar{2}\rangle$. Consequently, $|12\bar{2}\rangle$ was obtained as $|\psi(\theta_1)\rangle$ and $|1\bar{1}2\rangle$ was obtained as $|\psi(\theta_2)\rangle$ in ascending order of the cost function.

For condition (ii), the ansatz state $|\psi(\theta_1)\rangle$ was calculated to be $|1\bar{1}\bar{2}\rangle$ in all cases. Because $w_2 = 0$ in this case, no penalty was imposed on the deviation of the spin quantum number. In addition, the $|1\bar{1}\bar{2}\rangle$ state is orthogonal to $|1\bar{1}2\rangle$, resulting in no penalty being imposed by $\beta$. As a result, the cost function for $|1\bar{1}\bar{2}\rangle$ was the same as that for the ground state $|1\bar{1}2\rangle$. Therefore, $|1\bar{1}\bar{2}\rangle$ was consistently obtained as $|\psi(\theta_1)\rangle$ and other states were obtained as $|\psi(\theta_2)\rangle$. In the case of $w_1 < 37.4$, the charge neutral ground state $|1\bar{1}2\bar{2}\rangle$ with four electrons was obtained as $|\psi(\theta_2)\rangle$. Since the charge neutral ground state $|1\bar{1}2\bar{2}\rangle$ is orthogonal to the ground state of the ionized state $|1\bar{1}2\rangle$ and $w_2 = w_3 = 0$, only an $L_{penalty}$ of $(9/16)w_1$ was imposed on its cost function. Therefore, $\langle 1\bar{1}2\bar{2}|\hat{H}|1\bar{1}2\bar{2}\rangle + (9/16)w_1$ was less than the energy of the core ionized state $|12\bar{2}\rangle$ for $w_1 < 37.4$. The ordering of the $|1\bar{1}2\bar{2}\rangle$ and $|12\bar{2}\rangle$ cost functions was inverted in the case of $w_1 \geq 37.4$ and $|12\bar{2}\rangle$ was obtained as $|\psi(\theta_2)\rangle$.

The calculation results for condition (iii) were complicated. The calculations with $w_2 < 84.0$ provided the charge-neutral ground state $|1\bar{1}2\bar{2}\rangle$ as $|\psi(\theta_1)\rangle$ and the valence-double-ionized state $|1\bar{1}\rangle$ as $|\psi(\theta_2)\rangle$. Both states are orthogonal to the ground state $|1\bar{1}2\rangle$ and $w_1 = w_3 = 0$, such that an $L_{penalty}$ of $(1/4)w_2$ was imposed on $|1\bar{1}2\bar{2}\rangle$. However, for $w_2 < 84.0$, the $L_{penalty}$ value was less than the energy gap between $|1\bar{1}2\rangle$ and $|12\bar{2}\rangle$ and so $|1\bar{1}2\bar{2}\rangle$ was obtained as $|\psi(\theta_1)\rangle$. The same penalty was imposed on $|1\bar{1}\rangle$ for $w_1 = w_3 = 0$, and was less than the energy gap between $|1\bar{1}\rangle$ and $|12\bar{2}\rangle$. Consequently, $|1\bar{1}\rangle$ was obtained as $|\psi(\theta_2)\rangle$. In the case with $w_2 = 84.0$, the core-ionized state $|12\bar{2}\rangle$ was obtained as $|\psi(\theta_1)\rangle$ and $|1\bar{1}2\bar{2}\rangle$ was obtained as $|\psi(\theta_2)\rangle$ because $L(\theta_1) = \langle 12\bar{2}|\hat{H}|12\bar{2}\rangle$ was less than $L(\theta_2) = \langle 1\bar{1}2\bar{2}|\hat{H}|1\bar{1}2\bar{2}\rangle + (1/4)w_2$. In the case of $w_2 > 84.0$, the ground state $|1\bar{1}2\rangle$ was obtained as $|\psi(\theta_2)\rangle$ because $\langle 1\bar{1}2\bar{2}|\hat{H}|1\bar{1}2\bar{2}\rangle + (1/4)w_2$ became greater than $\langle 1\bar{1}2|\hat{H}|1\bar{1}2\rangle + \beta$.

The results obtained using condition (iv) were similar to those observed for condition (ii). The $|\psi(\theta_1)\rangle$ state was calculated to be $|1\bar{1}2\rangle$ in all cases and, although $|\Delta_2| = 1$, no penalty was imposed because $w_2 = 0$. In the case of $w_3 < 21.0$, the neutral ground state $|1\bar{1}2\bar{2}\rangle$ was obtained as $|\psi(\theta_2)\rangle$ because its $L_{penalty}$ of $w_3$ was less than the energy gap between $|1\bar{1}2\bar{2}\rangle$ and $|12\bar{2}\rangle$. The core-ionized state $|12\bar{2}\rangle$ was obtained as $|\psi(\theta_2)\rangle$ for $w_3 \geq 21.0$, suggesting that the ordering of the cost function was inverted due to the increase in $w_3$.

In conclusion, the weighting coefficients estimated from Eqs. (21) and (25) were adequate to raise the cost of undesired ansatz states so that the target state was obtained. In addition, the penalty terms imposed on the ansatz states shown in Table 5 suggest that fewer weighting coefficients were effective for this system. A penalty of $w_2$ was imposed on $|1\bar{1}2\rangle$, and so $w_2 = -\varepsilon_{1s}$ was sufficient to raise the cost function of $|1\bar{1}\bar{2}\rangle$ above that of $|12\bar{2}\rangle$. Accordingly, $(9/16)w_1 + w_3 > -(3/4)\varepsilon_{1s}$ was suitable for increasing the cost functions for $|1\bar{1}2\bar{2}\rangle$ and $|1\bar{1}\rangle$ such that they were higher than that for $|12\bar{2}\rangle$. The calculations using the weighting coefficients meeting these requirements were carried out as condition (v). The core-ionized state $|12\bar{2}\rangle$ and the neutral ground state $|1\bar{1}2\bar{2}\rangle$ were obtained as $|\psi(\theta_1)\rangle$ and $|\psi(\theta_2)\rangle$, respectively. Thus, the lower limit of each weighting coefficient was dependent on the system. The weighting coefficients estimated from Eqs. (21) and (25) were based on the minimum changes in the eigenvalues;[27] therefore, the estimated values were expected to be consistently equal to or larger than the lower limits that were required.

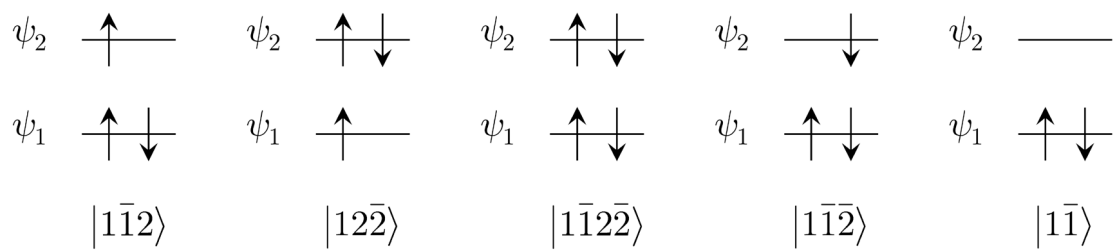

**Figure 1.** Primary configurations of the electronic states determined from VQD calculations of the core-ionized states for H₂O. Here, $\psi_1$ and $\psi_2$ correspond to the O 1s orbital and HOMO, respectively.

**Table 5.** Calculated energies, eigenvalues of the observables and imposed penalty terms of $|\psi_k\rangle$ states ($k \geq 1$) obtained from VQD calculations of the core-ionized states for H$_2$O. Each electronic state is denoted by a Slater determinant that corresponds to its main configuration as shown in Figure 1.

| Electronic state | Energy | Eigenvalue | | | Imposed penalty terms |
|---|---|---|---|---|---|
| | | $S(S+1)$ | $M_S$ | $N$ | |
| $|12\bar{2}\rangle$ | −55.4760 | 3/4 | 1/2 | 3 | 0 |
| $|1\bar{1}2\rangle$ | −75.5335 | 3/4 | 1/2 | 3 | $\beta$ |
| $|1\bar{1}2\bar{2}\rangle$ | −76.0268 | 0 | 0 | 4 | $(9/16)w_1 + (1/4)w_2 + w_3$ |
| $|1\bar{1}\bar{2}\rangle$ | −75.5335 | 3/4 | −1/2 | 3 | $w_2$ |
| $|1\bar{1}\rangle$ | −74.2798 | 0 | 0 | 2 | $(9/16)w_1 + (1/4)w_2 + w_3$ |

**Table 6.** Electronic state energies ($E(\theta_k)$) and assignments of $|\psi(\theta_k)\rangle$ for H$_2$O as calculated using the VQD method ($k$ = 0, 1 and 2). All values are in hartrees.

| Condition | $\beta$ | $w_2$ | $w_z$ | $w_N$ | $E(\theta_0)$ | Assignment | $E(\theta_1)$ | Assignment | $E(\theta_2)$ | Assignment |
|---|---|---|---|---|---|---|---|---|---|---|
| (i) | 15.0 | 37.4 | 84.0 | 21.0 | −75.5335 | $|1\bar{1}2\rangle$ | −75.5335 | $|1\bar{1}2\rangle$ | −55.4760 | $|12\bar{2}\rangle$ |
|  | 18.0 | 37.4 | 84.0 | 21.0 | −75.5335 | $|1\bar{1}2\rangle$ | −75.5335 | $|1\bar{1}2\rangle$ | −55.4760 | $|12\bar{2}\rangle$ |
|  | 21.0 | 37.4 | 84.0 | 21.0 | −75.5335 | $|1\bar{1}2\rangle$ | −55.4760 | $|12\bar{2}\rangle$ | −75.5335 | $|1\bar{1}2\rangle$ |
|  | 24.0 | 37.4 | 84.0 | 21.0 | −75.5335 | $|1\bar{1}2\rangle$ | −55.4760 | $|12\bar{2}\rangle$ | −75.5335 | $|1\bar{1}2\rangle$ |
|  | 27.0 | 37.4 | 84.0 | 21.0 | −75.5335 | $|1\bar{1}2\rangle$ | −55.4760 | $|12\bar{2}\rangle$ | −75.5335 | $|1\bar{1}2\rangle$ |
| (ii) | 21.0 | 24.0 | 0.0 | 0.0 | −75.5335 | $|1\bar{1}2\rangle$ | −75.5335 | $|1\bar{1}2\rangle$ | −76.0268 | $|1\bar{1}2\bar{2}\rangle$ |
|  | 21.0 | 32.0 | 0.0 | 0.0 | −75.5335 | $|1\bar{1}2\rangle$ | −75.5335 | $|1\bar{1}2\rangle$ | −76.0268 | $|1\bar{1}2\bar{2}\rangle$ |
|  | 21.0 | 37.4 | 0.0 | 0.0 | −75.5335 | $|1\bar{1}2\rangle$ | −75.5335 | $|1\bar{1}2\rangle$ | −55.4760 | $|12\bar{2}\rangle$ |
|  | 21.0 | 40.0 | 0.0 | 0.0 | −75.5335 | $|1\bar{1}2\rangle$ | −75.5335 | $|1\bar{1}2\rangle$ | −55.4760 | $|12\bar{2}\rangle$ |
|  | 21.0 | 48.0 | 0.0 | 0.0 | −75.5335 | $|1\bar{1}2\rangle$ | −75.5335 | $|1\bar{1}2\rangle$ | −55.4760 | $|12\bar{2}\rangle$ |
| (iii) | 21.0 | 0.0 | 60.0 | 0.0 | −75.5335 | $|1\bar{1}2\rangle$ | −76.0268 | $|1\bar{1}2\bar{2}\rangle$ | −74.2798 | $|1\bar{1}\rangle$ |
|  | 21.0 | 0.0 | 72.0 | 0.0 | −75.5335 | $|1\bar{1}2\rangle$ | −76.0268 | $|1\bar{1}2\bar{2}\rangle$ | −74.2798 | $|1\bar{1}\rangle$ |
|  | 21.0 | 0.0 | 84.0 | 0.0 | −75.5335 | $|1\bar{1}2\rangle$ | −55.4760 | $|12\bar{2}\rangle$ | −76.0268 | $|1\bar{1}2\bar{2}\rangle$ |
|  | 21.0 | 0.0 | 96.0 | 0.0 | −75.5335 | $|1\bar{1}2\rangle$ | −55.4760 | $|12\bar{2}\rangle$ | −75.5335 | $|1\bar{1}2\rangle$ |
|  | 21.0 | 0.0 | 108.0 | 0.0 | −75.5335 | $|1\bar{1}2\rangle$ | −55.4760 | $|12\bar{2}\rangle$ | −75.5335 | $|1\bar{1}2\rangle$ |
| (iv) | 21.0 | 0.0 | 0.0 | 15.0 | −75.5335 | $|1\bar{1}2\rangle$ | −75.5335 | $|1\bar{1}2\rangle$ | −76.0268 | $|1\bar{1}2\bar{2}\rangle$ |
|  | 21.0 | 0.0 | 0.0 | 18.0 | −75.5335 | $|1\bar{1}2\rangle$ | −75.5335 | $|1\bar{1}2\rangle$ | −76.0268 | $|1\bar{1}2\bar{2}\rangle$ |
|  | 21.0 | 0.0 | 0.0 | 21.0 | −75.5335 | $|1\bar{1}2\rangle$ | −75.5335 | $|1\bar{1}2\rangle$ | −55.4760 | $|12\bar{2}\rangle$ |
|  | 21.0 | 0.0 | 0.0 | 24.0 | −75.5335 | $|1\bar{1}2\rangle$ | −75.5335 | $|1\bar{1}2\rangle$ | −55.4760 | $|12\bar{2}\rangle$ |
|  | 21.0 | 0.0 | 0.0 | 27.0 | −75.5335 | $|1\bar{1}2\rangle$ | −75.5335 | $|1\bar{1}2\rangle$ | −55.4760 | $|12\bar{2}\rangle$ |
| (v) | 21.0 | 0.0 | 21.0 | 15.75 | −75.5335 | $|1\bar{1}2\rangle$ | −55.4760 | $|12\bar{2}\rangle$ | −76.0268 | $|1\bar{1}2\bar{2}\rangle$ |
|  | 21.0 | 7.0 | 21.0 | 11.8125 | −75.5335 | $|1\bar{1}2\rangle$ | −55.4760 | $|12\bar{2}\rangle$ | −76.0268 | $|1\bar{1}2\bar{2}\rangle$ |
|  | 21.0 | 14.0 | 21.0 | 7.875 | −75.5335 | $|1\bar{1}2\rangle$ | −55.4760 | $|12\bar{2}\rangle$ | −76.0268 | $|1\bar{1}2\bar{2}\rangle$ |
|  | 21.0 | 21.0 | 21.0 | 3.9375 | −75.5335 | $|1\bar{1}2\rangle$ | −55.4760 | $|12\bar{2}\rangle$ | −76.0268 | $|1\bar{1}2\bar{2}\rangle$ |
|  | 21.0 | 28.0 | 21.0 | 0.0 | −75.5335 | $|1\bar{1}2\rangle$ | −55.4760 | $|12\bar{2}\rangle$ | −76.0268 | $|1\bar{1}2\bar{2}\rangle$ |

**Table 7.** Calculated electronic state energies for $|\psi(\theta_k)\rangle$ ($E(\theta_k)$), cost functions ($L(\theta_k)$) and $L(\theta_k) - E(\theta_k)$ values ($k = 1, 2$). All values are in hartrees.

| Condition | $\beta$ | $w_1$ | $w_2$ | $w_3$ | $|\psi(\theta_1)\rangle$ | | | $|\psi(\theta_2)\rangle$ | | |
|---|---|---|---|---|---|---|---|---|---|---|
| | | | | | $E(\theta_1)$ | $L(\theta_1)$ | $L(\theta_1) - E(\theta_1)$ | $E(\theta_2)$ | $L(\theta_2)$ | $L(\theta_2) - E(\theta_2)$ |
| (i) | 15.0 | 37.4 | 84.0 | 21.0 | −75.5335 | −60.5335 | 15.0 | −55.4760 | −55.4760 | 0.0 |
| | 18.0 | 37.4 | 84.0 | 21.0 | −75.5335 | −57.5335 | 18.0 | −55.4760 | −55.4760 | 0.0 |
| | 21.0 | 37.4 | 84.0 | 21.0 | −55.4760 | −55.4760 | 0.0 | −75.5335 | −54.5335 | 21.0 |
| | 24.0 | 37.4 | 84.0 | 21.0 | −55.4760 | −55.4760 | 0.0 | −75.5335 | −51.5335 | 24.0 |
| | 27.0 | 37.4 | 84.0 | 21.0 | −55.4760 | −55.4760 | 0.0 | −75.5335 | −48.5335 | 27.0 |
| (ii) | 21.0 | 24.0 | 0.0 | 0.0 | −75.5335 | −75.5335 | 0.0 | −76.0268 | −62.5268 | 13.5 |
| | 21.0 | 32.0 | 0.0 | 0.0 | −75.5335 | −75.5335 | 0.0 | −76.0268 | −58.0268 | 18.0 |
| | 21.0 | 37.4 | 0.0 | 0.0 | −75.5335 | −75.5335 | 0.0 | −55.4760 | −55.4760 | 0.0 |
| | 21.0 | 40.0 | 0.0 | 0.0 | −75.5335 | −75.5335 | 0.0 | −55.4760 | −55.4760 | 0.0 |
| | 21.0 | 48.0 | 0.0 | 0.0 | −75.5335 | −75.5335 | 0.0 | −55.4760 | −55.4760 | 0.0 |
| (iii) | 21.0 | 0.0 | 60.0 | 0.0 | −76.0268 | −61.0268 | 15.0 | −74.2798 | −59.2798 | 15.0 |
| | 21.0 | 0.0 | 72.0 | 0.0 | −76.0268 | −58.0268 | 18.0 | −74.2798 | −56.2798 | 18.0 |
| | 21.0 | 0.0 | 84.0 | 0.0 | −55.4760 | −55.4760 | 0.0 | −76.0268 | −55.0268 | 21.0 |
| | 21.0 | 0.0 | 96.0 | 0.0 | −55.4760 | −55.4760 | 0.0 | −75.5335 | −54.5335 | 21.0 |
| | 21.0 | 0.0 | 108.0 | 0.0 | −55.4760 | −55.4760 | 0.0 | −75.5335 | −54.5335 | 21.0 |
| (iv) | 21.0 | 0.0 | 0.0 | 15.0 | −75.5335 | −75.5335 | 0.0 | −76.0268 | −61.0268 | 15.0 |
| | 21.0 | 0.0 | 0.0 | 18.0 | −75.5335 | −75.5335 | 0.0 | −76.0268 | −58.0268 | 18.0 |
| | 21.0 | 0.0 | 0.0 | 21.0 | −75.5335 | −75.5335 | 0.0 | −55.4760 | −55.4760 | 0.0 |
| | 21.0 | 0.0 | 0.0 | 24.0 | −75.5335 | −75.5335 | 0.0 | −55.4760 | −55.4760 | 0.0 |
| | 21.0 | 0.0 | 0.0 | 27.0 | −75.5335 | −75.5335 | 0.0 | −55.4760 | −55.4760 | 0.0 |
| (v) | 21.0 | 0.0 | 21.0 | 15.75 | −55.4760 | −55.4760 | 0.0 | −76.0268 | −55.0268 | 21.0 |
| | 21.0 | 7.0 | 21.0 | 11.8125 | −55.4760 | −55.4760 | 0.0 | −76.0268 | −55.0268 | 21.0 |
| | 21.0 | 14.0 | 21.0 | 7.875 | −55.4760 | −55.4760 | 0.0 | −76.0268 | −55.0268 | 21.0 |
| | 21.0 | 21.0 | 21.0 | 3.9375 | −55.4760 | −55.4760 | 0.0 | −76.0268 | −55.0268 | 21.0 |
| | 21.0 | 28.0 | 21.0 | 0.0 | −55.4760 | −55.4760 | 0.0 | −76.0268 | −55.0268 | 21.0 |

4.4. Calculation results for the TiO$_2$ and N-TiO$_2$ models

The optimized Cartesian coordinates for the models are summarized in Tables S1 and S2. The calculation results for the optimized structures are presented and discussed below.

4.4.1. HOMO → LUMO excited states

The HOMO and LUMO energy levels and weighting coefficients used in the calculations of HOMO → LUMO excited states are summarized in Table 8. The $\varepsilon_{\mathrm{HOMO-LUMO}}$ values were rounded up to the first decimal place prior to being used for the estimation of weighting coefficients. Since the HOMO → LUMO excitation energies were much smaller than the core-excitation energies and core-ionization energies, the weighting coefficients for the calculations of HOMO → LUMO excited states were also smaller.

The calculated excitation energies are presented in Table 9 along with the wavelengths converted from the excitation energies. The HOMO → LUMO excited states corresponding to the S$_1$ states were obtained as the $|\psi(\theta_1)\rangle$ ansatz states using the weighting coefficients in Table 8. The calculated absorption wavelength for Ti(OH)$_3$(NH$_2$) was longer than that for Ti(OH)$_4$, indicating that the absorption wavelength for TiO$_2$ should be redshifted following nitrogen doping, in qualitatively good agreement with experimental findings.[42,43] Figure 2 presents images of the HOMOs and LUMOs for the models as calculated using the Hartree-Fock method. The HOMO and LUMO for Ti(OH)$_4$ were mainly derived from O 2$p$ and Ti 3$d$ orbitals, respectively. In contrast, the HOMO for Ti(OH)$_3$(NH$_2$) originated from the N 2$p$ orbital. Because the energy level of the N 2$p$ orbital is higher than that of the O 2$p$ orbital (Table 8), the HOMO−LUMO gap is decreased by the substitution of an NH$_2$ group for an OH group, which is analogous to the redshift of the absorption wavelength for bulk TiO$_2$ as a result of nitrogen doping. The calculated HOMO → LUMO excitation energies were also greatly decreased following orbital optimization. The calculated energies for Ti(OH)$_4$ and Ti(OH)$_3$(NH$_2$) were 3.21 and 2.92 eV, respectively, in good agreement with the experimentally observed band gaps of 3.2 eV for TiO$_2$ and 2.5 eV for N-TiO$_2$.[42] These results suggest that the adopted models had similar electronic properties to those of the bulk systems, even though they were much smaller in size. It should also be noted that the calculation results obtained using a cluster model depend on the properties of the cluster and that dynamical correlations were not considered in the present VQD calculations. The errors derived from these conditions might counteract one another to provide successful results.

**Table 8.** The HOMO and LUMO energies obtained from Hartree-Fock calculations ($\varepsilon_{\text{HOMO}}$ and $\varepsilon_{\text{LUMO}}$), HOHO−LUMO gaps ($\varepsilon_{\text{HOMO−LUMO}}$), and weighting coefficients for the cost function in VQD calculations estimated from the $\varepsilon_{\text{HOMO−LUMO}}$ values, comprising overlap weights ($\beta$), s2 number weights ($w_1$), sz number weights ($w_2$) and particle number weights ($w_3$). All values are in hartrees.

| Model | $\varepsilon_{\text{HOMO}}$ | $\varepsilon_{\text{LUMO}}$ | $\varepsilon_{\text{HOMO−LUMO}}$ | Weighting coefficient | | | |
|---|---|---|---|---|---|---|---|
| | | | | $\beta$ | $w_1$ | $w_2$ | $w_3$ |
| Ti(OH)$_4$ | −0.4632 | 0.0463 | 0.5095 | 0.6 | 1.1 | 2.4 | 0.6 |
| Ti(OH)$_3$(NH$_2$) | −0.3876 | 0.0517 | 0.4394 | 0.5 | 0.9 | 2.0 | 0.5 |

**Table 9.** $S_0$ and $S_1$ electronic state energies and HOMO → LUMO excitation energies calculated for the Ti(OH)$_4$ and Ti(OH)$_3$(NH$_2$) models. Experimental values obtained for bulk TiO$_2$ and N-TiO$_2$ are also presented.

| Model | Orbital | Electronic state energy (hartrees) | | HOMO → LUMO excitation energy | | |
|---|---|---|---|---|---|---|
| | | $S_0$ | $S_1$ | (hartrees) | (eV) | (nm) |
| Ti(OH)$_4$ | not optimized | −1150.3944 | −1150.0545 | 0.3398 | 9.25 | 134.1 |
| | optimized | −1150.3944 | −1150.2764 | 0.1179 | 3.21 | 386.3 |
| TiO$_2$ (exptl.)[a] | | | | | 3.2 | 387.4 |
| Ti(OH)$_3$NH$_2$ | not optimized | −1130.5426 | −1130.2608 | 0.2817 | 7.67 | 161.7 |
| | optimized | −1130.5426 | −1130.4352 | 0.1073 | 2.92 | 424.5 |
| N-TiO$_2$ (exptl.)[a] | | | | | 2.5 | 495.9 |

[a] Data obtained from reference 42 based on the absorption band edge.

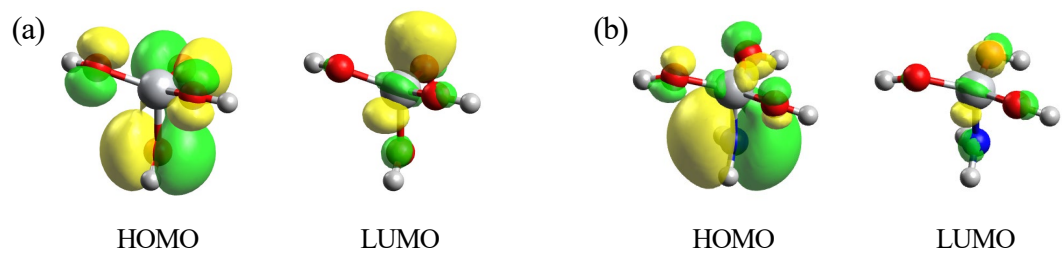

**Figure 2.** Images showing the HOMO and LUMO for (a) Ti(OH)$_4$ and (b) Ti(OH)$_3$(NH$_2$) as calculated using the Hartree-Fock method.

4.3. O 1s core-excited states

The O 1s and LUMO energy levels and weighting coefficients used in the calculations of the O 1s → LUMO excited states are summarized in Table 10. Note that the $\varepsilon_{1s-\mathrm{LUMO}}$ value was rounded up to the first decimal place prior to being used for the estimation of the weighting coefficients. The calculated O 1s → LUMO excited state energies for the Ti(OH)$_4$ and Ti(OH)$_3$(NH$_2$) models are provided in Table 11. These energies were averaged for the calculation of the core excitation energy value of each model, and the core excitation energy of Ti(OH)$_3$(NH$_2$) was found to be higher than that of Ti(OH)$_4$ by approximately 0.5 eV. These results suggest that the peak in the XANES spectrum of TiO$_2$ originating from the O 1s → LUMO excitation was shifted toward higher energy as a result of nitrogen doping, in agreement with experimental observations.[84] The core-excitation energies calculated with orbital optimization were generally in good agreement with the corresponding experimental values. The LUMOs shown in Figure 2 suggest that these orbitals primarily originated from Ti 3d orbitals and that their energy levels were shifted upward due to the substitution of a N atom for an O atom of Ti(OH)$_4$ (Table 8). In this oxide, the Ti−O bond is polarized to Ti$^{\delta+}$−O$^{\delta-}$ because of the high electronegativity of the oxygen atom.[85] Consequently, the electron density in the Ti atom is increased by the substitution of a N atom for an O atom because of the lower electronegativity of N compared with O. This effect, in turn, is responsible for the upward shift of the Ti 3d level.[71] As a result, the core excitation energy of Ti(OH)$_3$(NH$_2$) can be higher than that of Ti(OH)$_4$.

The calculated core-excitation energies were decreased by approximately 13 eV following orbital optimization, suggesting that significant orbital relaxation was associated with core-hole formation compared with HOMO-LUMO excitation. Although a hole in the HOMO primarily affects the electrons excited to the virtual orbitals, a core hole additionally affects the valence occupied orbitals, leading to significant relaxation. As a result, the calculated O 1s → LUMO core-excitation energies were greatly reduced through orbital optimization.

The calculated shift of the core-excitation energy due to the substitution of a N atom for an O atom was larger than the experimentally observed value. However, the substitution of a N atom for one of the O atoms in the Ti(OH)$_4$ model corresponded to a nitrogen doping level of 25 atom%, which was much larger than the concentration of less than 0.2 atom% applied in the experiments.[84] Therefore, the effects of nitrogen doping could be overestimated in the calculations results. A larger model capable of accurately simulating lower levels of doping would therefore be necessary for quantitatively accurate calculations.

**Table 10.** O 1s orbital energies calculated using the Hartree-Fock method ($\varepsilon_{1s}$) and the weighting coefficients used in the VQD calculations. All values are in hartrees.

| Model | $\varepsilon_{1s}$ | $\varepsilon_{\text{LUMO}}$ | $\varepsilon_{1s-\text{LUMO}}$ | Weighting coefficient | | | |
|---|---|---|---|---|---|---|---|
| | | | | $\beta$ | $w_1$ | $w_2$ | $w_3$ |
| Ti(OH)$_4$ | −20.5447 | 0.0463 | 20.5910 | 20.7 | 36.8 | 82.8 | 20.7 |
| | −20.5467 | | 20.5930 | | | | |
| | −20.5772 | | 20.6235 | | | | |
| | −20.5810 | | 20.6273 | | | | |
| Ti(OH)$_3$(NH$_2$) | −20.5397 | 0.0517 | 20.5914 | 20.7 | 36.8 | 82.8 | 20.7 |
| | −20.5624 | | 20.6141 | | | | |
| | −20.5635 | | 20.6152 | | | | |

**Table 11.** Calculated electronic energies for the ground states (S$_0$) and O 1$s$ → LUMO core-excited states and the calculated O 1$s$ → LUMO core-excitation energies.

| Model | Orbital | S$_0$ (hartrees) | O 1$s$ → LUMO core-excited state (hartrees) | Core-excitation energy (hartrees) | Core-excitation energy (eV) |
|---|---|---|---|---|---|
| Ti(OH)$_4$ | not optimized | −1150.3944 | −1129.9646 | 20.4298 | 555.92 |
|  | optimized | −1150.3944 | −1130.8285 | 19.5659 | 532.41 |
| TiO$_2$ (exptl.)[a] |  |  |  |  | 531.1 |
| Ti(OH)$_3$NH$_2$ | not optimized | −1130.5426 | −1110.0964 | 20.4462 | 556.36 |
|  | optimized | −1130.5426 | −1110.9545 | 19.5881 | 533.01 |
| N-TiO$_2$ (exptl.)[b] |  |  |  |  | 531.0 – 531.4 |

[a] Data taken from reference 84. [b] Data taken from reference 84. Note that the values were shifted upward with increasing amounts of nitrogen doping.

4.3. Ti 2p core-ionized states

Ti 2p orbital energies calculated using the Hartree-Fock method and the weighting coefficients estimated from these orbital energies are summarized in Table 12. The three orbital energies were determined to be approximately −18.08 hartree for Ti(OH)$_4$ and −18.05 hartree for Ti(OH)$_3$(NH$_2$). Note that all the $-\varepsilon_{2p}$ values, when rounded up to the first decimal place, were 18.1 hartree. As a result, common weighting coefficients were adopted for both models. The calculation results for the Ti 2p core-ionized states are provided in Table 13. The calculated core-ionization energies derived from the three Ti 2p orbitals were averaged to obtain the $E_{2p}$ value for each model and the resulting $E_{2p1/2}$ and $E_{2p3/2}$ values for Ti(OH)$_3$(NH$_2$) were found to be lower than those for Ti(OH)$_4$. The results suggest that the Ti 2p peaks in the XPS spectrum should be shifted to lower energy due to nitrogen doping, in agreement with the experimental findings.[70,71]

Since the Ti 2p level was shifted upward as a consequence of the substitution of a N atom for an O atom (Table 8), the Ti 2p core-ionization energy was decreased. However, the calculated core-ionized energies of Ti(OH)$_3$(NH$_2$) were lower than those of Ti(OH)$_4$ by 0.9 – 1.5 eV and this shift was larger than the experimental values of 0.3 – 0.6 eV.[70,71] This discrepancy can be attributed to the overestimated extent of nitrogen doping in the models, as with the discrepancy for the calculations of O 1s → LUMO core-excited states. The calculated core-ionization energies were decreased by approximately 15 eV following orbital optimization, suggesting that orbital relaxation had a significant effect.

Overall, the results of the VQD calculations suggest that nitrogen doping was responsible for the red-shift in the absorption band edge, the increase in the O 1s → Ti 3d excitation energy and the upward shift of the Ti 2p core-level, all of which were in qualitatively good agreement with the experimental data.

**Table 12.** Ti 2$p$ orbital energies calculated using the Hartree-Fock method ($\varepsilon_{2p}$) and the weighting coefficients used in the VQD calculations. All values are in hartrees.

| Model | $\varepsilon_{2p}$ | | | Weighting coefficient | | | |
|---|---|---|---|---|---|---|---|
| | Orbital 1 | Orbital 2 | Orbital 3 | $\beta$ | $w_1$ | $w_2$ | $w_3$ |
| Ti(OH)$_4$ | −18.0812 | −18.0815 | −18.0817 | 18.1 | 32.2 | 72.4 | 18.1 |
| Ti(OH)$_3$(NH$_2$) | −18.0461 | −18.0462 | −18.0465 | 18.1 | 32.2 | 72.4 | 18.1 |

**Table 13.** Calculated electronic energies of the charge-neutral ground states (S$_0$) and Ti 2$p$ core-ionized states, core-ionization energies ($E_{2p}$), and estimated $E_{2p1/2}$ and $E_{2p3/2}$ values. Corresponding experimental values are also presented.

| Model | Orbital | S$_0$ (hartrees)[a] | Ti 2$p$ core-ionized state (hartrees)[b] | $E_{2p}$ (hartrees) | $E_{2p}$ (eV) | $E_{2p1/2}$ (eV) | $E_{2p3/2}$ (eV) |
|---|---|---|---|---|---|---|---|
| Ti(OH)$_4$ | not optimized | −1150.3944 | −1132.3129 | 18.0815 | 492.01 | 495.81 | 490.11 |
|  | optimized | −1150.3944 | −1132.8492 | 18.0463 | 477.42 | 481.22 | 475.52 |
| Ti(OH)$_4$ (shifted)[c] |  |  |  |  | 460.5 | 464.3 | 458.6 |
| bulk TiO$_2$ (exptl.) |  |  |  |  | 460.5[d] | 464.3[e], 464.5[f] | 458.6[e], 458.7[f] |
| Ti(OH)$_3$NH$_2$ | not optimized | −1130.5426 | −1112.4963 | 17.5452 | 491.06 | 494.86 | 489.16 |
|  | optimized | −1130.5426 | −1113.0502 | 17.4924 | 475.98 | 479.78 | 474.08 |
| Ti(OH)$_3$NH$_2$ (shifted)[g] |  |  |  |  |  | 462.86 | 457.16 |
| bulk N-TiO$_2$ (exptl.) |  |  |  |  |  | 464.0,[h] 464.1[i] | 458.0,[e] 458.3[i] |

[a] Charge-neutral singlet ground state. [b] Averaged for each model. [c] Shifted downward by 16.92 eV from the calculated values after orbital optimizations so that the calculated $E_{2p}$ for Ti(OH)$_4$ coincides with the experimental value of 460.5 eV. [d] Estimated from the experimental $E_{2p1/2}$ and $E_{2p3/2}$ values in Reference 70 utilizing Eqs. (26) and (27). [e] Data taken from reference 70. [f] Data taken from reference 66. [g] Shifted downward by 16.92 eV from the calculated values for Ti(OH)$_3$(NH$_2$) after orbital optimizations. [h] Extracted from Figure 8(b) of reference 70. [i] Extracted from Figure 2(a) of reference 66.

## 5. Conclusions

The VQD method has attracted attention as a promising quantum-classical hybrid algorithm for the calculation of excited states. In the present study, calculations of the core-excited states and core-ionized states for common molecules utilizing the VQD method were simulated on a classical computer, focusing on the penalty terms in the cost function. The weighting coefficients for these penalty terms were estimated based on molecular orbital levels and the minimum deviations of the eigenvalues of the wavefunction regarding the spin and the number of electrons. Adopting this simple procedure allowed the core-level states to be successfully calculated. The relationship between the weighting coefficients and the resulting ansatz states was systematically examined based on calculations of the O 1$s$ core-ionized state for a water molecule. The results indicate that the weighting coefficients estimated by our procedure allowed the target states to be obtained by raising the cost of undesired states. The O 1$s$ core-excited states and the Ti 2$p$ core-ionized states for $TiO_2$ and N-$TiO_2$ photocatalysts serving as model compounds could also be calculated in the same manner as the valence excited states. The results of such calculations demonstrated that nitrogen doping induced a red-shift in the absorption band edge, increased the O 1$s$ → Ti 3$d$ excitation energy and raised the Ti 2$p$ core-level. These results were consistent with experimental findings, suggesting that VQD calculations can be applied to the analysis of functional materials in collaboration with experimental work. The results of this study are expected to provide useful guidelines for further applications of the VQD method and also to inspire further development of the algorithm.

**References**

1.	Cao, Y.; Romero, J.; Olson, J. P.; Degroote, M.; Johnson, P. D.; Kieferová, M.; Kivlichan, I. D.; Menke, T.; Peropadre, B.; Sawaya, N. P. D.; Sim, S.; Veis, L.; Aspuru-Guzik, A., Quantum chemistry in the age of quantum computing. *Chem. Rev.* **2019,** *119* (19), 10856-10915.
2.	McArdle, S.; Endo, S.; Aspuru-Guzik, A.; Benjamin, S. C.; Yuan, X., Quantum computational chemistry. *Rev. Mod. Phys.* **2020,** *92* (1), 015003.

3.      Sherrill, C. D.; Schaefer, H. F., The configuration interaction method: advances in highly correlated approaches. *Adv. Quantum Chem.* **1999,** *34*, 143-269.

4.      Olsen, J.; Jørgensen, P.; Koch, H.; Balkova, A.; Bartlett, R. J., Full configuration–interaction and state of the art correlation calculations on water in a valence double-zeta basis with polarization functions. *J. Chem. Phys.* **1996,** *104* (20), 8007-8015.

5.      Abrams, D. S.; Lloyd, S., Quantum algorithm providing exponential speed increase for finding eigenvalues and eigenvectors. *Phys. Rev. Lett.* **1999,** *83* (24), 5162-5165.

6.      Aspuru-Guzik, A.; Dutoi Anthony, D.; Love Peter, J.; Head-Gordon, M., Simulated quantum computation of molecular energies. *Science* **2005,** *309* (5741), 1704-1707.

7.      Wecker, D.; Bauer, B.; Clark, B. K.; Hastings, M. B.; Troyer, M., Gate-count estimates for performing quantum chemistry on small quantum computers. *Phys. Rev. A* **2014,** *90* (2), 022305.

8.      Babbush, R.; Berry, D. W.; Sanders, Y. R.; Kivlichan, I. D.; Scherer, A.; Wei, A. Y.; Love, P. J.; Aspuru-Guzik, A., Exponentially more precise quantum simulation of fermions in the configuration interaction representation. *Quantum Sci. Technol.* **2017,** *3* (1), 015006.

9.      Preskill, J., Quantum computing in the NISQ era and beyond. *Quantum* **2018,** *2*, 79-98.

10.     Wang, D.; Higgott, O.; Brierley, S., Accelerated variational quantum eigensolver. *Phys. Rev. Lett.* **2019,** *122* (14), 140504.

11.     Sugisaki, K.; Sakai, C.; Toyota, K.; Sato, K.; Shiomi, D.; Takui, T., Bayesian phase difference estimation: a general quantum algorithm for the direct calculation of energy gaps. *Phys. Chem. Chem. Phys.* **2021,** *23* (36), 20152-20162.

12.     Peruzzo, A.; McClean, J.; Shadbolt, P.; Yung, M.-H.; Zhou, X.-Q.; Love, P. J.; Aspuru-Guzik, A.; O'Brien, J. L., A variational eigenvalue solver on a photonic quantum processor. *Nat. Commun.* **2014,** *5* (1), 4213-4219.

13.     McClean, J. R.; Romero, J.; Babbush, R.; Aspuru-Guzik, A., The theory of variational hybrid quantum-classical algorithms. *New J. Phys.* **2016,** *18* (2), 023023-1 - 023023-22.

14.     Colless, J. I.; Ramasesh, V. V.; Dahlen, D.; Blok, M. S.; Kimchi-Schwartz, M. E.; McClean, J. R.; Carter, J.; de Jong, W. A.; Siddiqi, I., Computation of molecular spectra on a quantum processor with an error-resilient algorithm. *Phys. Rev. X.* **2018,** *8* (1).

15.     Parrish, R. M.; Hohenstein, E. G.; McMahon, P. L.; Martínez, T. J., Quantum computation of electronic transitions using a variational quantum eigensolver. *Phys. Rev. Lett.* **2019,** *122* (23), 230401-1 - 230401-6.

16.     Nakanishi, K. M.; Mitarai, K.; Fujii, K., Subspace-search variational quantum eigensolver for excited states. *Phys. Rev. Res.* **2019,** *1* (3), 033062-1 - 033062-7.

17.     Ollitrault, P. J.; Kandala, A.; Chen, C. F.; Barkoutsos, P. K.; Mezzacapo, A.; Pistoia, M.; Sheldon, S.; Woerner, S.; Gambetta, J. M.; Tavernelli, I., Quantum equation of motion for computing molecular excitation energies on a noisy quantum processor. *Phys. Rev. Res.* **2020,** *2* (4), 043140-1 - 043140-13.

18. Higgott, O.; Wang, D.; Brierley, S., Variational quantum computation of excited states. *Quantum* **2019,** *3*, 156-166.

19. Ibe, Y.; Nakagawa, Y. O.; Earnest, N.; Yamamoto, T.; Mitarai, K.; Gao, Q.; Kobayashi, T., Calculating transition amplitudes by variational quantum deflation. *arXiv preprint arXiv:2002.11724* **2020**.

20. Gao, Q.; Jones, G. O.; Motta, M.; Sugawara, M.; Watanabe, H. C.; Kobayashi, T.; Watanabe, E.; Ohnishi, Y.-y.; Nakamura, H.; Yamamoto, N., Applications of quantum computing for investigations of electronic transitions in phenylsulfonyl-carbazole TADF emitters. *Npj Comput. Mater.* **2021,** *7* (1), 70-78.

21. Shuai, Z.; Peng, Q., Organic light-emitting diodes: theoretical understanding of highly efficient materials and development of computational methodology. *Natl. Sci. Rev.* **2017,** *4* (2), 224-239.

22. Prentice, A. W.; Zwijnenburg, M. A., The role of computational chemistry in discovering and understanding organic photocatalysts for renewable fuel synthesis. *Adv. Energy Mater.* **2021,** *11* (29), 2100709.

23. Cui, Y.; Zhu, P.; Liao, X.; Chen, Y., Recent advances of computational chemistry in organic solar cell research. *J. Mater. Chem. C* **2020,** *8* (45), 15920-15939.

24. Irie, M.; Fukaminato, T.; Matsuda, K.; Kobatake, S., Photochromism of diarylethene molecules and crystals: memories, switches, and actuators. *Chem. Rev.* **2014,** *114* (24), 12174-12277.

25. Ryabinkin, I. G.; Genin, S. N.; Izmaylov, A. F., Constrained variational quantum eigensolver: quantum computer search engine in the Fock space. *J. Chem. Theory Comput.* **2019,** *15* (1), 249-255.

26. Greene-Diniz, G.; Ramo, D. M., Generalized unitary coupled cluster excitations for multireference molecular states optimized by the variational quantum eigensolver. *Int. J. Quantum Chem.* **2021,** *121* (4), e26352.

27. Kuroiwa, K.; Nakagawa, Y. O., Penalty methods for a variational quantum eigensolver. *Phys. Rev. Res.* **2021,** *3* (1), 013197-1-013197-11.

28. Penner-Hahn, J. E., X-ray absorption spectroscopy in coordination chemistry. *Coord. Chem. Rev.* **1999,** *190-192*, 1101-1123.

29. Penner-Hahn, J. E., 2.13 - X-ray Absorption Spectroscopy. In *Comprehensive Coordination Chemistry II*, McCleverty, J. A.; Meyer, T. J., Eds. Pergamon: Oxford, 2003; pp 159-186.

30. Henderson, G. S.; de Groot, F. M. F.; Moulton, B. J. A., X-ray absorption near-edge structure (XANES) spectroscopy. *Rev. Mineral. Geochem.* **2014,** *78* (1), 75-138.

31. Chen, Y.; Chen, C.; Zheng, C.; Dwaraknath, S.; Horton, M. K.; Cabana, J.; Rehr, J.; Vinson, J.; Dozier, A.; Kas, J. J.; Persson, K. A.; Ong, S. P., Database of ab initio L-edge X-ray absorption near edge structure. *Sci. Data* **2021,** *8* (1), 153.

32. Fadley, C. S., X-ray photoelectron spectroscopy: progress and perspectives. *J. Electron. Spectrosc. Relat. Phenom.* **2010,** *178-179*, 2-32.

33. Starr, D. E.; Liu, Z.; Hävecker, M.; Knop-Gericke, A.; Bluhm, H., Investigation of solid/vapor interfaces using ambient pressure X-ray photoelectron spectroscopy. *Chem. Soc. Rev.* **2013,** *42* (13), 5833-5857.

34. Brundle, C. R.; Crist, B. V., X-ray photoelectron spectroscopy: a perspective on quantitation accuracy for composition analysis of homogeneous materials. *J. Vac. Sci. Technol.* **2020,** *38* (4), 041001.

35. Greczynski, G.; Hultman, L., X-ray photoelectron spectroscopy: towards reliable binding energy referencing. *Prog. Mater Sci.* **2020,** *107*, 100591.

36. Norman, P.; Dreuw, A., Simulating X-ray spectroscopies and calculating core-excited states of molecules. *Chem. Rev.* **2018,** *118* (15), 7208-7248.

37. Bagus, P. S.; Ilton, E. S.; Nelin, C. J., The interpretation of XPS spectra: insights into materials properties. *Surf. Sci. Rep.* **2013,** *68* (2), 273-304.

38. Fujishima, A.; Honda, K., Electrochemical photolysis of water at a semiconductor electrode. *Nature* **1972,** *238* (5358), 37-38.

39. Fox, M. A.; Dulay, M. T., Heterogeneous photocatalysis. *Chem. Rev.* **1993,** *93* (1), 341-357.

40. Hoffmann, M. R.; Martin, S. T.; Choi, W.; Bahnemann, D. W., Environmental applications of semiconductor photocatalysis. *Chem. Rev.* **1995,** *95* (1), 69-96.

41. Hashimoto, K.; Irie, H.; Fujishima, A., $TiO_2$ photocatalysis: a historical overview and future prospects. *Jpn. J. Appl. Phys.* **2005,** *44* (12), 8269-8285.

42. Asahi, R.; Morikawa, T.; Ohwaki, T.; Aoki, K.; Taga, Y., Visible-light photocatalysis in nitrogen-doped titanium oxides. *Science* **2001,** *293* (5528), 269-271.

43. Asahi, R.; Morikawa, T.; Irie, H.; Ohwaki, T., Nitrogen-doped titanium dioxide as visible-light-sensitive photocatalyst: Designs, developments, and prospects. *Chem. Rev.* **2014,** *114* (19), 9824-9852.

44. Shirai, S.; Yamamoto, S.; Hyodo, S.-a., Accurate calculation of core-electron binding energies: Multireference perturbation treatment. *J. Chem. Phys.* **2004,** *121* (16), 7586-7594.

45. Bauman, N. P.; Liu, H.; Bylaska, E. J.; Krishnamoorthy, S.; Low, G. H.; Granade, C. E.; Wiebe, N.; Baker, N. A.; Peng, B.; Roetteler, M.; Troyer, M.; Kowalski, K., Toward quantum computing for high-energy excited states in molecular systems: quantum phase estimations of core-level states. *J. Chem. Theory Comput.* **2021,** *17* (1), 201-210.

46. Baerends, E. J.; Gritsenko, O. V.; van Meer, R., The Kohn–Sham gap, the fundamental gap and the optical gap: the physical meaning of occupied and virtual Kohn–Sham orbital energies. *Phys. Chem. Chem. Phys.* **2013,** *15* (39), 16408-16425.

47. Koopmans, T., Über die zuordnung von wellenfunktionen und eigenwerten zu den einzelnen elektronen eines atoms. *Physica* **1934,** *1* (1), 104-113.

48. Raghavachari, K.; Trucks, G. W.; Pople, J. A.; Head-Gordon, M., A fifth-order perturbation comparison of electron correlation theories. *Chem. Phys. Lett.* **1989,** *157* (6), 479-483.

49. Watts, J. D.; Gauss, J.; Bartlett, R. J., Coupled-cluster methods with noniterative triple excitations for restricted open-shell Hartree–Fock and other general single determinant reference functions. Energies and analytical gradients. *J. Chem. Phys.* **1993,** *98* (11), 8718-8733.

50. Kucharski, S. A.; Bartlett, R. J., Noniterative energy corrections through fifth-order to the coupled cluster singles and doubles method. *J. Chem. Phys.* **1998,** *108* (13), 5243-5254.

51. Becke, A. D., Density-functional thermochemistry. III. The role of exact exchange. *J. Chem. Phys.* **1993,** *98* (7), 5648-5652.

52. Stephens, P. J.; Devlin, F. J.; Chabalowski, C. F.; Frisch, M. J., Ab initio calculation of vibrational absorption and circular dichroism spectra using density functional force fields. *J. Phys. Chem.* **1994,** *98* (45), 11623-11627.

53. Gaussian 09, Revision D.01, Frisch, M. J.; Trucks, G. W; Schlegel, H. B.; Scuseria, G. E.; Robb, M. A.; Cheeseman, J. R.; Scalmani, G.; Barone, V.; Mennucci, B.; Petersson, G. A.; Nakatsuji, H. et al., Gaussian, Inc., Wallingford CT, 2016.

54. QunaSys Inc., Qamuy, https://qunasys.com/services/qamuy.

55. Kandala, A.; Mezzacapo, A.; Temme, K.; Takita, M.; Brink, M.; Chow, J. M.; Gambetta, J. M., Hardware-efficient variational quantum eigensolver for small molecules and quantum magnets. *Nature* **2017,** *549* (7671), 242-246.

56. Mishra, S. K.; Ram, B., Quasi-Newton methods. In *Introduction to Unconstrained Optimization, R*, Mishra, S. K.; Ram, B., Eds. Springer Singapore: Singapore, 2019; pp 245-289.

57. Jordan, P.; Wigner, E., Über das paulische äquivalenzverbot. *Z. Phys.* **1928,** *47* (9), 631-651.

58. Ortiz, G.; Gubernatis, J. E.; Knill, E.; Laflamme, R., Quantum algorithms for fermionic simulations. *Phys. Rev. A* **2001,** *64* (2), 022319-1 - 022319-14.

59. Roos, B. O.; Taylor, P. R.; Sigbahn, P. E. M., A complete active space SCF method (CASSCF) using a density matrix formulated super-CI approach. *Chem. Phys.* **1980,** *48* (2), 157-173.

60. Roos, B. O., The complete active space self-consistent field method and its applications in electronic structure calculations. *Adv. Chem. Phys.* **1987,** *69*, 399-445.

61. Levine, B. G.; Durden, A. S.; Esch, M. P.; Liang, F.; Shu, Y., CAS without SCF—Why to use CASCI and where to get the orbitals. *J. Chem. Phys.* **2021,** *154* (9), 090902.

62. Dunning, T. H., Gaussian basis sets for use in correlated molecular calculations. I. The atoms boron through neon and hydrogen. *J. Chem. Phys.* **1989,** *90* (2), 1007-1023.

63. Balabanov, N. B.; Peterson, K. A., Systematically convergent basis sets for transition metals.

I. All-electron correlation consistent basis sets for the 3d elements Sc–Zn. *J. Chem. Phys.* **2005,** *123* (6), 064107.

64. Balabanov, N. B.; Peterson, K. A., Basis set limit electronic excitation energies, ionization potentials, and electron affinities for the 3d transition metal atoms: Coupled cluster and multireference methods. *J. Chem. Phys.* **2006,** *125* (7), 074110.

65. Fernando, A.; Weerawardene, K. L. D. M.; Karimova, N. V.; Aikens, C. M., Quantum mechanical studies of large metal, metal oxide, and metal chalcogenide nanoparticles and clusters. *Chem. Rev.* **2015,** *115* (12), 6112-6216.

66. Lu, X.; Xu, X.; Wang, N. Q.; Zhang, Q.; Ehara, M.; Nakatsuji, H., Cluster modeling of metal oxides: how to cut out a cluster? *Chem. Phys. Lett.* **1998,** *291* (3-4), 445-452.

67. Persson, P.; Gebhardt, J. C. M.; Lunell, S., The smallest possible nanocrystals of semiionic oxides. *J. Phys. Chem. B* **2003,** *107* (15), 3336-3339.

68. Band, Y. B.; Avishai, Y., 4 - Spin. In *Quantum Mechanics with Applications to Nanotechnology and Information Science*, Band, Y. B.; Avishai, Y., Eds. Academic Press: Amsterdam, 2013; pp 159-192.

69. Biesinger, M. C.; Lau, L. W. M.; Gerson, A. R.; Smart, R. S. C., Resolving surface chemical states in XPS analysis of first row transition metals, oxides and hydroxides: Sc, Ti, V, Cu and Zn. *Appl. Surf. Sci.* **2010,** *257* (3), 887-898.

70. Zhang, Z.; Goodall, J. B. M.; Morgan, D. J.; Brown, S.; Clark, R. J. H.; Knowles, J. C.; Mordan, N. J.; Evans, J. R. G.; Carley, A. F.; Bowker, M.; Darr, J. A., Photocatalytic activities of N-doped nano-titanias and titanium nitride. *J. Eur. Ceram. Soc.* **2009,** *29* (11), 2343-2353.

71. Jia, T.; Fu, F.; Yu, D.; Cao, J.; Sun, G., Facile synthesis and characterization of N-doped $TiO_2$/C nanocomposites with enhanced visible-light photocatalytic performance. *Appl. Surf. Sci.* **2018,** *430*, 438-447.

72. X-ray Photoelectron Spectroscopy. In *Fundamentals of Nanoscale Film Analysis*, Alford, T. L.; Feldman, L. C.; Mayer, J. W., Eds. Springer US: Boston, MA, 2007; pp 199-213.

73. Ozaki, T.; Lee, C.-C., Absolute binding energies of core levels in solids from first principles. *Phys. Rev. Lett.* **2017,** *118* (2), 026401.

74. Kahk, J. M.; Lischner, J., Accurate absolute core-electron binding energies of molecules, solids, and surfaces from first-principles calculations. *Phys. Rev. Mater.* **2019,** *3* (10), 100801.

75. Logsdail, A. J.; Scanlon, D. O.; Catlow, C. R. A.; Sokol, A. A., Bulk ionization potentials and band alignments from three-dimensional periodic calculations as demonstrated on rocksalt oxides. *Phys. Rev. B* **2014,** *90* (15), 155106.

76. Yoshida, H.; Yamada, K.; Tsutsumi, J.; Sato, N., Complete description of ionization energy and electron affinity in organic solids: Determining contributions from electronic polarization, energy band dispersion, and molecular orientation. *Phys. Rev. B* **2015,** *92* (7), 075145.

77. Jiang, H.; Shen, Y.-C., Ionization potentials of semiconductors from first-principles. *J. Chem. Phys.* **2013,** *139* (16), 164114.

78. Domke, M.; Xue, C.; Puschmann, A.; Mandel, T.; Hudson, E.; Shirley, D. A.; Kaindl, G., Carbon and oxygen K-edge photoionization of the CO molecule. *Chem. Phys. Lett.* **1990,** *173* (1), 122-128.

79. Hitchcock, A. P.; Brion, C. E., K-shell excitation spectra of CO, $N_2$ and $O_2$. *J. Electron. Spectrosc. Relat. Phenom.* **1980,** *18* (1), 1-21.

80. Remmers, G.; Domke, M.; Puschmann, A.; Mandel, T.; Xue, C.; Kaindl, G.; Hudson, E.; Shirley, D. A., High-resolution K-shell photoabsorption in formaldehyde. *Phys. Rev. A* **1992,** *46* (7), 3935-3944.

81. Hitchcock, A. P.; Brion, C. E., K-shell excitation of HCN by electron energy loss spectroscopy. *J. Electron. Spectrosc. Relat. Phenom.* **1979,** *15* (1), 201-206.

82. Jolly, W. L.; Bomben, K. D.; Eyermann, C. J., Core-electron binding energies for gaseous atoms and molecules. *At. Data Nucl. Data Tables* **1984,** *31* (3), 433-493.

83. Bagus, P. S., Self-consistent-field wave functions for hole states of some Ne-like and Ar-like ions. *Phys. Rev.* **1965,** *139* (3A), A619-A634.

84. Stewart, S. J.; Fernández-García, M.; Belver, C.; Mun, B. S.; Requejo, F. G., Influence of N-doping on the structure and electronic properties of titania nanoparticle photocatalysts. *J. Phys. Chem. B* **2006,** *110* (33), 16482-16486.

85. Allred, A. L., Electronegativity values from thermochemical data. *J. Inorg. Nucl. Chem.* **1961,** *17* (3), 215-221.

# Supplementary Information

# Calculation of core-excited and core-ionized states using variational quantum deflation method and applications to photocatalyst modelling

Soichi Shirai,[†][*] Takahiro Horiba[†] and Hirotoshi Hirai[†]

[†] Toyota Central R&D Laboratories, Inc., Nagakute, Aichi 480−1192, Japan

[*] Author to whom correspondence should be addressed.

E−mail: shirai@mosk.tytlabs.co.jp

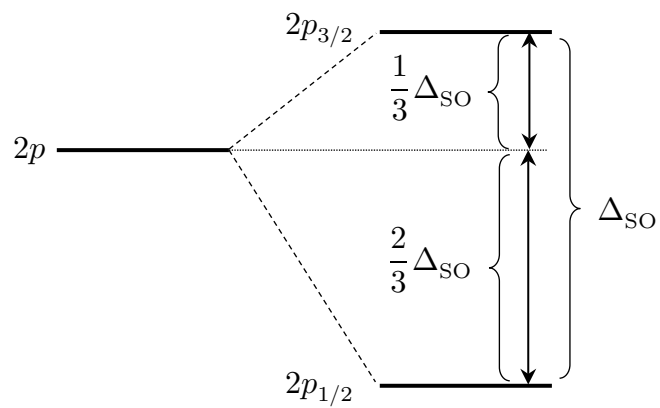

**Figure S1.** Splitting of the 2p core-ionized doublet state of an atom into $2p_{1/2}$ and $2p_{3/2}$ states due to spin-orbit interactions.

**Table S1.** Cartesian coordinates of the (TiO$_2$)$_{16}$ cluster for an optimized geometry. All coordinates are in Å.

| Atom number | Atom | x | y | z |
|---|---|---|---|---|
| 1 | O | −2.743094 | −4.930878 | −1.644764 |
| 2 | O | −3.223842 | 2.620283 | −1.794587 |
| 3 | TI | −1.491642 | −3.525124 | −1.651573 |
| 4 | O | −2.349298 | −2.078782 | −1.656494 |
| 5 | O | −0.446784 | 1.520713 | −2.765294 |
| 6 | TI | −1.351217 | 2.643908 | −1.833069 |
| 7 | O | −0.575438 | 4.286053 | −1.952746 |
| 8 | O | 0.320667 | −3.429621 | −2.149929 |
| 9 | TI | 1.600307 | −2.205244 | −1.819556 |
| 10 | O | 1.428093 | −0.742141 | −2.786001 |
| 11 | TI | 1.506863 | 1.104128 | −2.384429 |
| 12 | O | 1.734851 | 3.058321 | −2.985297 |
| 13 | TI | 1.280138 | 4.393438 | −2.035112 |
| 14 | O | 3.389917 | −2.548544 | −1.454051 |
| 15 | O | 3.352293 | 1.232853 | −2.009813 |
| 16 | O | 2.706991 | 5.549568 | −1.612746 |
| 17 | TI | −3.297590 | −4.721143 | 0.077714 |
| 18 | O | −4.174841 | −3.194449 | 0.205574 |
| 19 | TI | −3.234491 | −1.501819 | 0.249338 |
| 20 | O | −3.991495 | 0.079553 | −0.349306 |
| 21 | TI | −3.502224 | 1.792519 | −0.216095 |
| 22 | O | −1.588480 | −3.789473 | 0.254565 |
| 23 | O | −1.365606 | −1.161963 | 0.649103 |
| 24 [a] | O | −1.572274 | 1.808945 | −0.062399 |
| 25 | O | 1.572274 | −1.808945 | 0.062399 |
| 26 | O | 1.365606 | 1.161963 | −0.649103 |
| 27 | O | 1.588480 | 3.789473 | −0.254565 |
| 28 | TI | 3.502224 | −1.792519 | 0.216095 |
| 29 | O | 3.991495 | −0.079553 | 0.349306 |
| 30 | TI | 3.234491 | 1.501819 | −0.249338 |
| 31 | O | 4.174841 | 3.194449 | −0.205574 |
| 32 | TI | 3.297590 | 4.721143 | −0.077714 |
| 33 | O | −2.706991 | −5.549568 | 1.612746 |
| 34 | O | −3.352293 | −1.232853 | 2.009813 |
| 35 [a] | O | −3.389917 | 2.548544 | 1.454051 |
| 36 | TI | −1.280138 | −4.393438 | 2.035112 |
| 37 | O | −1.734851 | −3.058321 | 2.985297 |
| 38 | TI | −1.506863 | −1.104128 | 2.384429 |
| 39 [a] | O | −1.428093 | 0.742141 | 2.786001 |
| 40 [a] | TI | −1.600307 | 2.205244 | 1.819556 |
| 41 [a] | O | −0.320667 | 3.429621 | 2.149929 |
| 42 | O | 0.575438 | −4.286053 | 1.952746 |
| 43 | TI | 1.351217 | −2.643908 | 1.833069 |
| 44 | O | 0.446784 | −1.520713 | 2.765294 |
| 45 | O | 2.349298 | 2.078782 | 1.656494 |
| 46 | TI | 1.491642 | 3.525124 | 1.651573 |
| 47 | O | 3.223842 | −2.620283 | 1.794587 |
| 48 | O | 2.743094 | 4.930878 | 1.644764 |

[a] Cut out from the (TiO$_2$)$_{16}$ cluster to prepare the Ti(OH)$_4$ model.

**Table S2.** Cartesian coordinates of the Ti(OH)$_4$ and Ti(OH)$_3$(NH$_2$) cluster models for the optimized geometries. All coordinates are in Å.

| Model | Atom number | Atom | x | y | z |
|---|---|---|---|---|---|
| Ti(OH)$_4$ | 1 | TI | 0.143200 | 0.068143 | −0.098753 |
| | 2 | O | −0.968933 | −0.801362 | 1.207634 |
| | 3 | O | −1.387677 | −0.113054 | −1.137972 |
| | 4 | O | 0.421414 | 1.757011 | 0.317694 |
| | 5 | O | 1.663469 | −0.885107 | −0.258325 |
| | 6 | H | 1.850555 | −1.784273 | −0.561348 |
| | 7 | H | −0.531650 | −1.073469 | 2.032125 |
| | 8 | H | −0.249525 | 2.438470 | 0.472430 |
| | 9 | H | −2.045952 | −0.739775 | −0.802893 |
| Ti(OH)$_3$(NH$_2$) | 1 | TI | 0.163467 | 0.043089 | −0.101975 |
| | 2 | N | −1.058976 | −0.576716 | 1.247628 |
| | 3 | O | −1.365352 | −0.046302 | −1.155998 |
| | 4 | O | 0.633124 | 1.717598 | 0.179693 |
| | 5 | O | 1.562620 | −1.090119 | −0.147826 |
| | 6 | H | 1.592837 | −2.055431 | −0.193166 |
| | 7 | H | −0.782548 | −0.775609 | 2.212358 |
| | 8 | H | 0.180257 | 2.471373 | 0.579572 |
| | 9 | H | −1.781404 | −0.906971 | −1.313207 |
| | 10 | H | −2.035728 | −0.293731 | 1.217554 |